\documentclass{aa}

\usepackage{graphicx}
\usepackage{txfonts}

\usepackage{makecell}
\usepackage{multirow}

\usepackage{hyperref}
\hypersetup{
    colorlinks=true,
    linkcolor=blue,
    citecolor=blue,
    filecolor=magenta,      
    urlcolor=blue,
}

\usepackage{microtype}
\begin{document}

%%%%%%%%%%%%%%%%%%%%%%%%%%%%%%%%%%%%%%%%%%%%%%%%%%%%%%%%%%%%%%%

   \title{Composition gradients in sub-Neptunes}

   \subtitle{K2-18\,b and TOI-270\,d as case studies}

%%%%%%%%%%%%%%%%%%%%%%%%%%%%%%%%%%%%%%%%%%%%%%%%%%%%%%%%%%%%%%%

\author{Luca Morf\corrauth{luca.morf@uzh.ch} \and Ravit Helled}

\institute{Department of Astrophysics, University of Zurich, Winterthurerstrasse 190, 8057 Zurich, Switzerland}

\date{Received 26 June 2026, Accepted 29 October 2026}

%%%%%%%%%%%%%%%%%%%%%%%%%%%%%%%%%%%%%%%%%%%%%%%%%%%%%%%%%%%%%%

\abstract{
%Context.
Structure models of sub-Neptunes commonly assume purely adiabatic interiors with distinct layers of homogeneous composition. 
}
{
%Aims.
We assess how allowing for more complex interiors with composition gradients affects the inferred internal structure and bulk compositions of the sub-Neptunes K2-18\,b and TOI-270\,d.
}
{
%Methods.
We compare purely adiabatic models with distinct layers against models that include composition gradients and stable non-convective regions.
We present solutions that are consistent with the observed masses, radii, and atmospheric boundary conditions.
}
{
%Results.
Generally, interior degeneracy remains substantial even at a fixed mass and radius.
We further find that composition gradients significantly increase the range of viable interior structures.
For example, the maximum hydrogen--helium (H--He) mass fraction can increase significantly, by a factor of about three at a 1\,bar temperature of 300\,K and by a factor of about 10 at 600\,K for K2-18\,b. 
For TOI-270\,d, the increase is generally more than an order of magnitude.
Composition gradients additionally weaken parameter correlations. 
For K2-18\,b and TOI-270\,d, the Spearman rank correlation between the H--He and iron abundances drops from $\sim0.7$ to $\sim0.3$, with a similar weakening between the H--He abundance and the rock-to-iron mass ratio.
However, other correlations, such as that between the H--He abundance and the water-to-rock mass ratio, persist or even strengthen and remain non-negligible.
}
{
%Conclusions.
Purely adiabatic models underestimate the range of plausible compositions and overestimate the impact of more precise measurements, atmosphere models, and host-star constraints on a more accurate characterisation. 
We suggest that interior models of sub-Neptunes should by default include more complex interiors, such as composition gradients and non-convective regions, when using data for interpreting the planetary structure, formation, and evolution.
}

%%%%%%%%%%%%%%%%%%%%%%%%%%%%%%%%%%%%%%%%%%%%%%%%%%%%%%%%%%%%%%


%%%%%%%%%%%%%%%%%%%%%%%%%%%%%%%%%%%%%%%%%%%%%%%%%%%%%%%%%%%%%%

\maketitle
\nolinenumbers

%%%%%%%%%%%%%%%%%%%%%%%%%%%%%%%%%%%%%%%%%%%%%%%%%%%%%%%%%%%%%%

\section{Introduction}

%%%%%%%%%%%%%%%%%%%%%%%%%%%%%%%%%%%%%%%%%%%%%%%%%%%%%%%%%%%%%%

The discovery of exoplanets has fundamentally changed our understanding of planetary systems. 
Since the first detection \citep{Mayor1995}, the number of known planets beyond the Solar System has grown rapidly, revealing a wide diversity of planetary types and system architectures \citep[e.g.][]{Otegi2020a, Parc2024}. 
Among these, a prominent class consists of planets with sizes and masses between those of Earth and Neptune, commonly referred to as super-Earths or sub-Neptunes. 
As no direct analogues to these planets exist in the Solar System, they provide important constraints for planet formation theories, which must explain how such planets can form efficiently and frequently \citep[e.g.][]{Venturini2020}.
In addition, it has been suggested that some of these planets may host conditions favourable for life, for instance through the presence of extended surface water oceans \citep[e.g.][]{MolLous2022, Madhusudhan2023}. 

%%%%%%%%%%%%%%%%%%%%%%%%%%%%%%%%%%%%%%%%%%%%%%%%%%%%%%%%%%%%%%

In this work, we investigate the internal structures of sub-Neptunes using two representative case studies.
Although the choice of targets is arbitrary, we focus on K2-18\,b \citep{Montet2015} and TOI-270\,d \citep{Gunther2019}, as they were investigated in various studies using different approaches \citep[e.g.][]{Madhusudhan2020, Benneke2024, Howard2025, Schmidt2025, Rigby2026}.
Also, as both planets have been proposed to be potentially habitable, a detailed understanding of their interiors is desirable.
Here, we focus on inferring their interior composition and structure.
This task is fundamentally challenging, as the internal structure of planets cannot be observed directly. 

%%%%%%%%%%%%%%%%%%%%%%%%%%%%%%%%%%%%%%%%%%%%%%%%%%%%%%%%%%%%%%

\begin{figure}
    \centering
    \includegraphics[width=\linewidth]{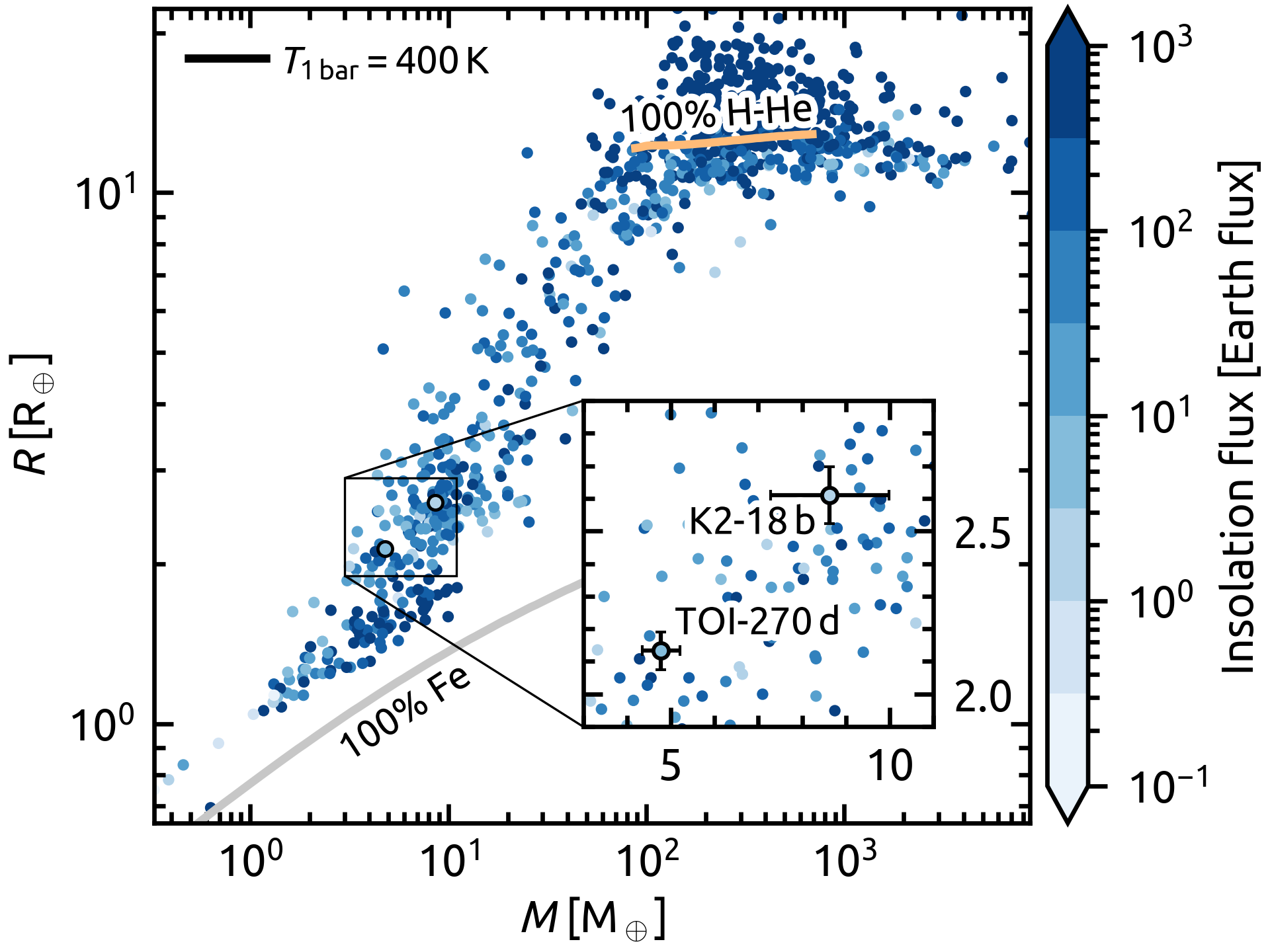}
    \caption{
    Masses, radii, and insolation fluxes of exoplanets in the DACE catalogue.
    The two planets used as case studies in this work are highlighted with their measurement uncertainties. 
    Also shown are lines corresponding to pure H--He and pure iron (Fe) adiabatic models. 
    }
    \label{fig:MR}
\end{figure}

Figure~\ref{fig:MR} shows the mass--radius distribution ($M$--$R$ relation) of exoplanets in the Data Analysis Center for Exoplanets (DACE) catalogue \citep{Otegi2020a, Parc2024}.
The planets selected as case studies in this work are highlighted, together with their observational uncertainties.
For reference, we also include two limiting categories of composition models representing idealised end-members. 
The first corresponds to planets composed entirely of a hydrogen--helium (H--He) mixture \citep[according to][]{Chabrier2021} with protosolar composition (see Sect.~\ref{sec:Methods}), while the second represents pure iron planets \citep[according to][]{Attia2026}. 
Both models assume a purely adiabatic temperature profile.
These two cases are intended to bracket the expected range of the planetary bulk compositions, since they represent the lightest and heaviest abundant materials to form planetary interiors. 
Most observed exoplanets should lie between these two limits.
Some deviations are expected, particularly for strongly irradiated close-in planets or young planets with elevated internal temperatures that lead to inflated radii.
Such planets are in particular incompatible with the assumed 1\,bar temperature of 400\,K in Fig.~\ref{fig:MR}.

%%%%%%%%%%%%%%%%%%%%%%%%%%%%%%%%%%%%%%%%%%%%%%%%%%%%%%%%%%%%%%

%%%%%%%%%%%%%%%%%%%%%%%%%%%%%%%%%%%%%%%%%%%%%%%%%%%%%%%%%%%%%%

\begin{table*}
\caption{
Physical data for the two planets used as case studies in this work.
}
\centering
    \begin{tabular}{lllllll}
    \hline
    \hline
     & K2-18\,b & & TOI-270\,d & \\
    \hline
    \hline
    $M$ [M$_\oplus$] & 8.63$\,\pm\,$1.35 & \cite{Cloutier2019} & 4.78$\,\pm\,$0.43 & \cite{VanEylen2021} \\
    \hline
    $R$ [R$_\oplus$] & 2.610$\,\pm\,$0.087 & \cite{Benneke2019} & 2.133$\,\pm\,$0.058 & \cite{VanEylen2021} \\
    \hline
    $T_{1\,\mathrm{bar}}^\mathrm{low}$ [K] & 300 & \makecell[l]{\cite{Benneke2019} \\ \cite{Madhusudhan2023} \\ \cite{Schmidt2025}} & 400 & \makecell[l]{\cite{Constantinou2026} \\ \cite{Rigby2026}} \\
    \hline
    $T_{1\,\mathrm{bar}}^\mathrm{high}$ [K] & 600 & \makecell[l]{\cite{Blain2021} \\ \cite{Wogan2024} \\ \cite{Schmidt2025}} & 700 & \makecell[l]{\cite{Benneke2024} \\ \cite{Nixon2025} \\ \cite{Constantinou2026} \\ \cite{Rigby2026}} \\
    \hline
    \hline
    \end{tabular}
\tablefoot{
Since the cited literature often gives a range of possible 1\,bar temperatures, we selected two values, $T_{1\,\mathrm{bar}}^\mathrm{low}$ and $T_{1\,\mathrm{bar}}^\mathrm{high}$, for each planet that are representative of the lower and higher ends found in the literature. 
}
\label{tab:planet_data}
\end{table*}

%%%%%%%%%%%%%%%%%%%%%%%%%%%%%%%%%%%%%%%%%%%%%%%%%%%%%%%%%%%%%%

The two planets considered as case studies in this work lie within the broad region between the H--He-dominated and iron-dominated regimes.
This region is intrinsically degenerate, as notably different internal compositions can reproduce the same observed mass and radius.
In an extreme example, both a nearly pure water world and a planet consisting of an iron core overlain by a H--He envelope may occupy the same position in the mass--radius diagram.
Consequently, mass and radius measurements alone provide remarkably limited constraints on interior structure within this region of parameter space \citep{Otegi2020b}.

%%%%%%%%%%%%%%%%%%%%%%%%%%%%%%%%%%%%%%%%%%%%%%%%%%%%%%%%%%%%%%

Furthermore, constraints from interior structure models are also limited by the assumptions built into the models.
Simplified, purely adiabatic interiors with homogeneous and distinct layers can fail to provide a complete overview of the possible parameter space permitted by data. 
The planets in the Solar System provide a clear example of this problem. 
Since high-precision gravity field data are available in the Solar System as an additional constraint \citep[e.g.][]{Wirth2026}, interior models must include more complex structures.
For Jupiter, models assuming perfectly mixed and homogeneous layers with adiabatic temperature profiles have proven insufficient to match the full set of observations \citep[e.g.][]{Wahl2017, Militzer2024}. 
More flexible structures, including composition gradients, a fuzzy core, and partial mixing are required. 
Similar results are found in the case of Saturn \citep[e.g.][]{Mankovich2021}.  
Such structures are an expected outcome of the formation process \citep[e.g.][]{Helled2017, Stevenson2022, Valletta2022} and have also been successfully applied to Uranus and Neptune \citep[e.g.][]{Neuenschwander2024, Morf2024, TejadaArevalo2025, Morf2025}.

%%%%%%%%%%%%%%%%%%%%%%%%%%%%%%%%%%%%%%%%%%%%%%%%%%%%%%%%%%%%%%

The existence of composition gradients implies that the planetary interior consists of regions that are stable against convection and therefore non-adiabatic. 
Evolution models clearly show that these features influence thermal evolution, planetary radius, and atmospheric composition \citep[e.g.][]{Vazan2020, Knierim2024, TejadaArevalo2025, Eberlein2025, Eberlein2026}.
The lesson from the Solar System is hence clear: 
a good understanding of the nature of planetary interiors requires a diversity of models that includes more complex features, such as composition gradients and non-convective regions. 

%%%%%%%%%%%%%%%%%%%%%%%%%%%%%%%%%%%%%%%%%%%%%%%%%%%%%%%%%%%%%%

In this study, we investigate the influence of allowing for such features by comparing them to simple purely adiabatic models for the two sub-Neptunes selected as case studies.
Our paper is organised as follows.
Section~\ref{sec:Methods} introduces our methods for modelling the internal structure as well as the various assumptions adopted in the model. 
In Sect.~\ref{sec:Results} we present the inferred interior models for K2-18\,b.
Analogous results for TOI-270\,d can be found in Appendix~\ref{sec:further_figures}.
Finally, we discuss potential limitations in Sect.~\ref{sec:Limitations} and summarise our conclusions in Sect.~\ref{sec:Conclusions}.

%%%%%%%%%%%%%%%%%%%%%%%%%%%%%%%%%%%%%%%%%%%%%%%%%%%%%%%%%%%%%%

\section{Methods}
\label{sec:Methods}

%%%%%%%%%%%%%%%%%%%%%%%%%%%%%%%%%%%%%%%%%%%%%%%%%%%%%%%%%%%%%%

The data used by the interior structure models consist of the mass, radius, and temperature at the 1\,bar pressure level (see Table~\ref{tab:planet_data}).
Regarding the interiors, we assume that they consist of a mixture of H--He \citep{Chabrier2021}, water \citep{CanoAmoros2026}, forsterite (Mg$_2$SiO$_4$; hereafter referred to as rock)\footnote{As provided by \cite{TejadaArevalo2026} under \url{https://doi.org/10.5281/zenodo.21812109}.} \citep{Stewart2020}, and iron \citep{Attia2026}.
We now present two descriptions of the internal structure: purely adiabatic models and composition-gradient models.
The purely adiabatic models represent an idealised and maximally simple configuration. 
In this case, each planet is divided into several distinct, homogeneous layers. 
Each layer contains a single material. 
The layers are ordered by increasing density, i.e. from an outer H--He envelope, with a protosolar mass ratio of 0.705 to 0.275, to a rock or iron core in the deepest interior. 
Each layer is hydrostatically stable (see Eq.~\eqref{eq:HE}) and the temperature gradient is assumed to be adiabatic.
Heat transport is assumed to occur exclusively by convection.
At the interfaces between adjacent layers, a Ledoux-stable jump in temperature is enforced (see Eq.~\eqref{eq:grad_T_stable}).
As our models are static and do not solve the energy transport equation, the mechanism of heat transport across these composition interfaces is not treated explicitly.
Convection is expected to be inhibited at such interfaces where heat is then transported by radiative diffusion or conduction.

%%%%%%%%%%%%%%%%%%%%%%%%%%%%%%%%%%%%%%%%%%%%%%%%%%%%%%%%%%%%%%

In contrast, composition-gradient models generally allow multiple materials to coexist within the same layer.
Furthermore, they permit, but do not require, regions where composition varies continuously with radius and which are stable against convection.
Therefore, both stable and convective regions, in which the composition is homogenised by efficient mixing, can be present within the same planet for composition-gradient models.
For this case, we used a simplified version of the algorithm developed by \cite{Morf2024, Morf2025}, designed to minimise prior assumptions and to provide an agnostic assessment of the interior structure and bulk composition. 
This approach demonstrates that even the interiors of Uranus and Neptune (for which we have gravity data measurements) are not as constrained as traditionally thought, allowing for both rock-dominated and ice-dominated solutions. 
We refer the reader to the original publications for full details and now provide a summary of the methods employed in this work.

%%%%%%%%%%%%%%%%%%%%%%%%%%%%%%%%%%%%%%%%%%%%%%%%%%%%%%%%%%%%%%

To construct the internal structure, we began from random density profiles, $\rho(r)$, where $r$ denotes the radial coordinate.
The resulting gravitational field, $U(r)$, was then computed using the theory of figures\footnote{\url{https://doi.org/10.5281/zenodo.16902935} provides the numerical implementation.} \citep{Zharkov1978}. 
This allows the pressure profile, $P(r)$, to be inferred under the assumption of hydrostatic equilibrium,
\begin{equation}
    \vec{\nabla} P = \rho \vec{\nabla} U.
    \label{eq:HE}
\end{equation}
The temperature profile $T(r)$ and the composition profile $\vec{X}(r)$ were derived subsequently. 
This step is both degenerate and difficult. 
It is degenerate because density and pressure alone do not uniquely determine temperature and composition, especially when several materials coexist. 
It is difficult because a density profile generated without reference to material properties is unlikely to satisfy physical constraints, such as consistency with an adiabatic temperature gradient, $\nabla_T = \nabla_{\mathrm{ad}}$.
To address these issues, we used an iterative scheme \citep[see also][]{Morf2025} in which the density profile is progressively adjusted until a physically consistent solution is obtained. 
Convergence was reached when the pressure profile no longer varied significantly between iterations. 
Specifically, we adopted a relative pressure deviation threshold of $\varepsilon = 0.02$ that quantifies convergence.
All quantities, including density, pressure, temperature, and composition, were discretised on a grid with $N = 512$ points. 
Finally, note that we ensured that hydrogen and water are always miscible according to \cite{Howard2025}. 
This ensures that, under the given planetary temperature and pressure conditions, no such demixing occurs and the interior structure remains unchanged. 
An analysis of our results regarding the miscibility between hydrogen and rock is presented in Appendix~\ref{sec:hydro_rock_miscibility}.

%%%%%%%%%%%%%%%%%%%%%%%%%%%%%%%%%%%%%%%%%%%%%%%%%%%%%%%%%%%%%%

We did not adopt the full method of \cite{Morf2025}, because the available observational constraints are limited. 
Three main simplifications were introduced. 
First, the planets were assumed to be non-rotating. 
This implies perfect spherical symmetry and vanishing gravitational moments.
In particular, one could hence also use a 1D integration to infer $U(r)$ rather than the theory of figures.
Second, the composition was always assumed to either remain constant or to vary linearly as a function of radius $r$. 
Third, in regions with composition gradients, the temperature gradient was prescribed directly by Eq.~\eqref{eq:grad_T_stable} rather than generated randomly. 
We assumed
\begin{equation}
\nabla_T = \nabla_{\mathrm{ad}} + R_\rho B,
\label{eq:grad_T_stable}
\end{equation}
where 
\begin{align}
    B &= \frac{\chi_\rho}{\chi_T} \left(\frac{\mathrm{d} \ln \rho}{\mathrm{d} \ln \vec{X}}\right)_{P,T} \nabla_{\vec{X}}, 
    &\quad \chi_\rho &= \left(\frac{\mathrm{d} \ln P}{\mathrm{d} \ln \rho}\right)_{T,\vec{X}}, \nonumber \\
    \nabla_{\vec{X}} &= \frac{\mathrm{d} \ln \vec{X}}{\mathrm{d} \ln P}, 
    &\quad \chi_T &= \left(\frac{\mathrm{d} \ln P}{\mathrm{d} \ln T}\right)_{\rho,\vec{X}}.
\end{align}
Motivated by \cite{Howard2025}, we adopted a constant value of $R_\rho = 0.01$.
Note that $\vec{X}(r)=\left[X(r), Y(r), Z_1(r), \dots\right]$, where $X(r), Y(r), Z_1(r), \dots$ are the mass fractions as a function of $r$ for hydrogen, helium, and heavier constituents, respectively.

%%%%%%%%%%%%%%%%%%%%%%%%%%%%%%%%%%%%%%%%%%%%%%%%%%%%%%%%%%%%%%

Since the mean molecular weight always increases as $r$ decreases, we have $B>0$ in our models; Eq.~\eqref{eq:grad_T_stable} therefore satisfies the Ledoux stability criterion \citep{Ledoux_1947},
\begin{equation}
\nabla_T < \nabla_{\mathrm{ad}} + B,
\label{eq:ledoux_stability}
\end{equation}
by construction.
However, the Ledoux stability criterion does not exclude double-diffusive convection \citep{Leconte2012}.
\cite{French2019} suggest that double-diffusive convection in ice-giant interiors requires values of $R_\rho$ comparable to unity.
In that regard, our adopted value of $R_\rho=0.01$ provides a conservative choice.
Even $R_\rho=0.1$ would remain well below the range in which double-diffusive convection is expected.
To test the sensitivity of our results to the choice of $R_\rho$, we present further results with $R_\rho=0.1$ in Appendix~\ref{sec:sens_test_Rp}.

%%%%%%%%%%%%%%%%%%%%%%%%%%%%%%%%%%%%%%%%%%%%%%%%%%%%%%%%%%%%%%

The three simplifications introduced above substantially reduce the number of free parameters that can be varied when generating $\rho(r)$.
We provide an in-depth discussion of our parameter-space sampling procedure, including the adopted priors and comparisons between the prior and posterior distributions, in Appendix~\ref{sec:sampling_methods}.
Models whose total H--He mass fraction is below $10^{-6}$ are not shown in the figures but are included in all Spearman correlation coefficients.
In most cases, we randomly generated model parameters from their respective prior distributions and retained only those models that satisfied the observational constraints.
When this direct random sampling was inefficient because the allowed region of parameter space was small, we employed a Markov chain Monte Carlo (MCMC) approach to sample the reduced parameter space more efficiently \citep{ForemanMackey2013}.

%%%%%%%%%%%%%%%%%%%%%%%%%%%%%%%%%%%%%%%%%%%%%%%%%%%%%%%%%%%%%%

\begin{figure*}
    \centering
    \includegraphics[width=0.49\linewidth]{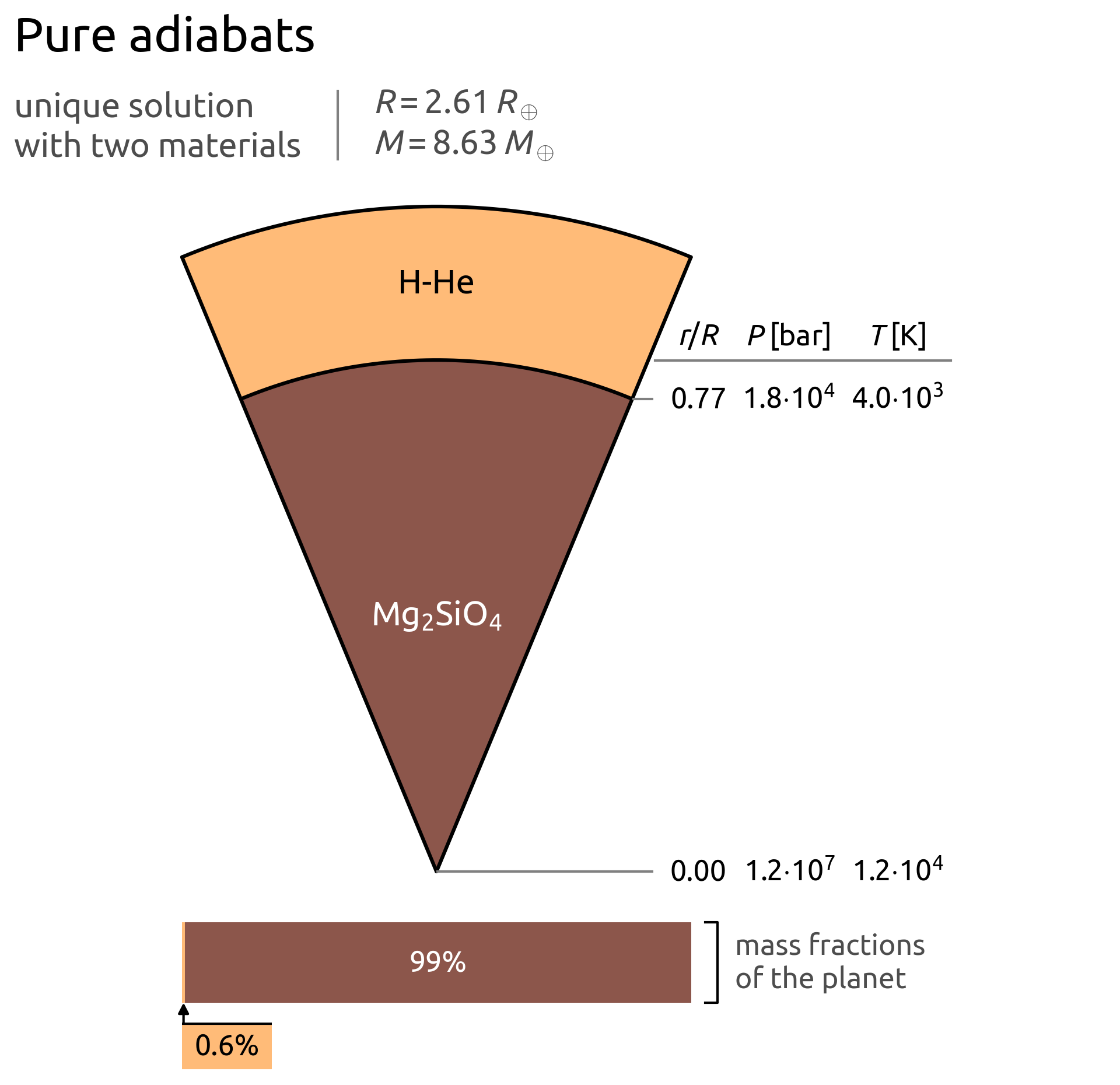}
    \includegraphics[width=0.49\linewidth]{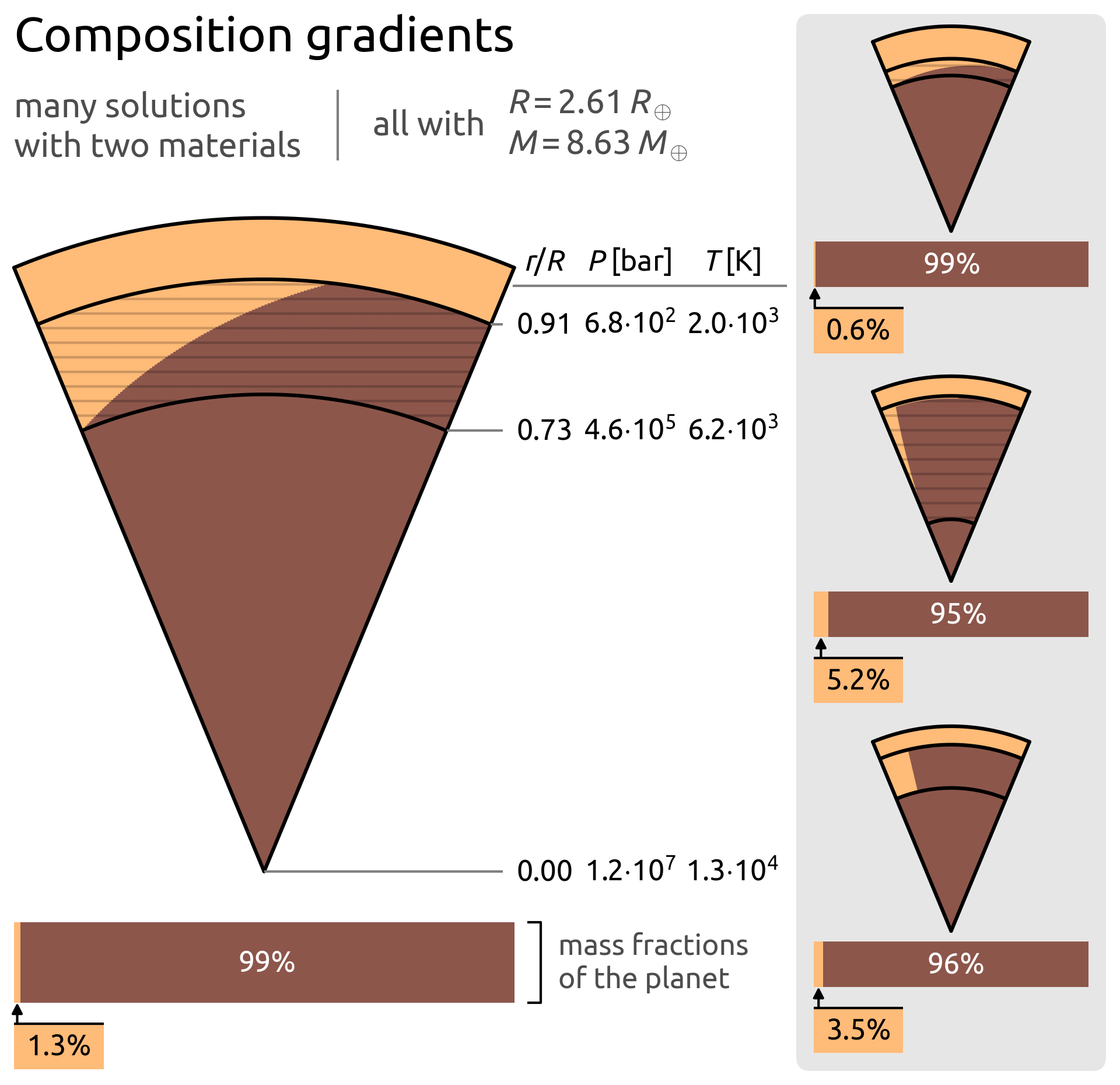}
    \caption{
    Examples of interior models for planet K2-18\,b. 
    Multicoloured arcs show relative mass fractions of the different materials.   
    For example, if H--He and Mg$_2$SiO$_4$ each have a mass fraction of 0.5 at $r/R=0.85$, the arc at $r/R=0.85$ consists of two halves with different colours. 
    Left: Purely adiabatic model with two single-material layers. 
    Since only two materials are used, the solution is unique for the given radius $R$ and mass $M$. 
    Right: Solutions when allowing for composition gradients, where the composition varies linearly in stable regions (dashed).  
    The percentage numbers indicate the bulk abundances. 
    The adopted radius $R$ and mass $M$ are identical for all the solutions. 
    }
    \label{fig:K2-18b_HHe-rock_slice_pure+grad-s0-T300}
\end{figure*}

%%%%%%%%%%%%%%%%%%%%%%%%%%%%%%%%%%%%%%%%%%%%%%%%%%%%%%%%%%%%%%

Finally, we note that the literature infers the adopted 1\,bar temperatures (Table~\ref{tab:planet_data}) with an atmosphere model, usually containing a metallicity factor relative to solar values.
Nevertheless, we did not impose a constraint on the atmospheric metallicity in our models and rather left it as a free parameter.
This is because the atmospheric metallicity does not necessarily represent the composition of the outermost layers \citep[see e.g.][]{Muller2024, Muller2026}.
In addition, atmospheric measurements cannot be easily translated into atmospheric heavy-element mass fractions as we discuss in detail in Appendix~\ref{sec:atmos_metal}.
Still, in this appendix we also present results for models that account for atmospheric metallicity constraints to show that our main conclusions persist in such a case.

%%%%%%%%%%%%%%%%%%%%%%%%%%%%%%%%%%%%%%%%%%%%%%%%%%%%%%%%%%%%%%

In the following, we investigate both the purely adiabatic and the composition-gradient cases by considering additional subcategories.
First, we distinguish between models that incorporate the full observational uncertainties, $\sigma_{R},\sigma_{M}>0$ (up to $2\sigma_\mathrm{observation}$ to account for 95\% of solutions), and models in which the uncertainties are assumed to be negligible, $\sigma_{R},\sigma_{M}=0$ (in practice $0.01\sigma_\mathrm{observation}$).
Second, we adopted two different values for the 1\,bar temperature ($T_{1\,\mathrm{bar}}^\mathrm{low}$ and $T_{1\,\mathrm{bar}}^\mathrm{high}$) as outer boundary conditions (see Table~\ref{tab:planet_data}).
Both 1\,bar temperature cases are motivated by observations and represent low- and high-temperature end-members to explore the range of interior structures compatible with the observations.
High temperatures may be maintained over gigayear timescales through inefficient cooling associated with composition gradients or boundary layers \citep[e.g.][and references therein]{Vazan2020, Eberlein2025, Eberlein2026}.

%%%%%%%%%%%%%%%%%%%%%%%%%%%%%%%%%%%%%%%%%%%%%%%%%%%%%%%%%%%%%%

\section{The composition of sub-Neptunes}
\label{sec:Results}

%%%%%%%%%%%%%%%%%%%%%%%%%%%%%%%%%%%%%%%%%%%%%%%%%%%%%%%%%%%%%%

We investigate how different physical assumptions and external constraints affect the inferred interior compositions of K2-18\,b and TOI-270\,d. 
We first consider simplified two-material models and examine the origin of degeneracy in the cases of a purely adiabatic interior and one with composition gradients (Sect.~\ref{sec:Results_2_materials}).
We then extend the analysis to four-material models, providing a larger range of plausible compositions (Sect.~\ref{sec:Results_4_materials}).
Next, we assess the extent to which atmospheric boundary conditions and host-star abundance measurements can reduce interior degeneracy (Sect.~\ref{sec:Results_atmosphere_hoststar}). 
Finally, we focus on the inferred water-to-rock mass ratio, H$_2$O/Mg$_2$SiO$_4$ (hereafter the ice-to-rock ratio), and further correlations between the different materials (Sect.~\ref{sec:Results_ice_rock_correlations}). 

%%%%%%%%%%%%%%%%%%%%%%%%%%%%%%%%%%%%%%%%%%%%%%%%%%%%%%%%%%%%%%

We note that all figures in this section are shown for K2-18\,b. 
Equivalent figures for TOI-270\,d are provided in Appendix~\ref{sec:further_figures}. 
These supplementary figures demonstrate that the qualitative conclusions presented here do not correspond to a specific planet but are valid in general. 

%%%%%%%%%%%%%%%%%%%%%%%%%%%%%%%%%%%%%%%%%%%%%%%%%%%%%%%%%%%%%%

\subsection{Inferred composition using two materials}
\label{sec:Results_2_materials}

%%%%%%%%%%%%%%%%%%%%%%%%%%%%%%%%%%%%%%%%%%%%%%%%%%%%%%%%%%%%%%
 
We begin by discussing interior models that contain only two materials: a H--He mixture and rock (Mg$_2$SiO$_4$). 
Although no known planet is expected to consist exclusively of these two materials, this simplified setup provides a useful baseline.  
It also resembles idealised interior models that have been considered in the literature \citep[e.g.][]{Tian2024, Rogers2025}. 

%%%%%%%%%%%%%%%%%%%%%%%%%%%%%%%%%%%%%%%%%%%%%%%%%%%%%%%%%%%%%%

\begin{figure*}
    \centering
    \includegraphics[width=0.49\linewidth]{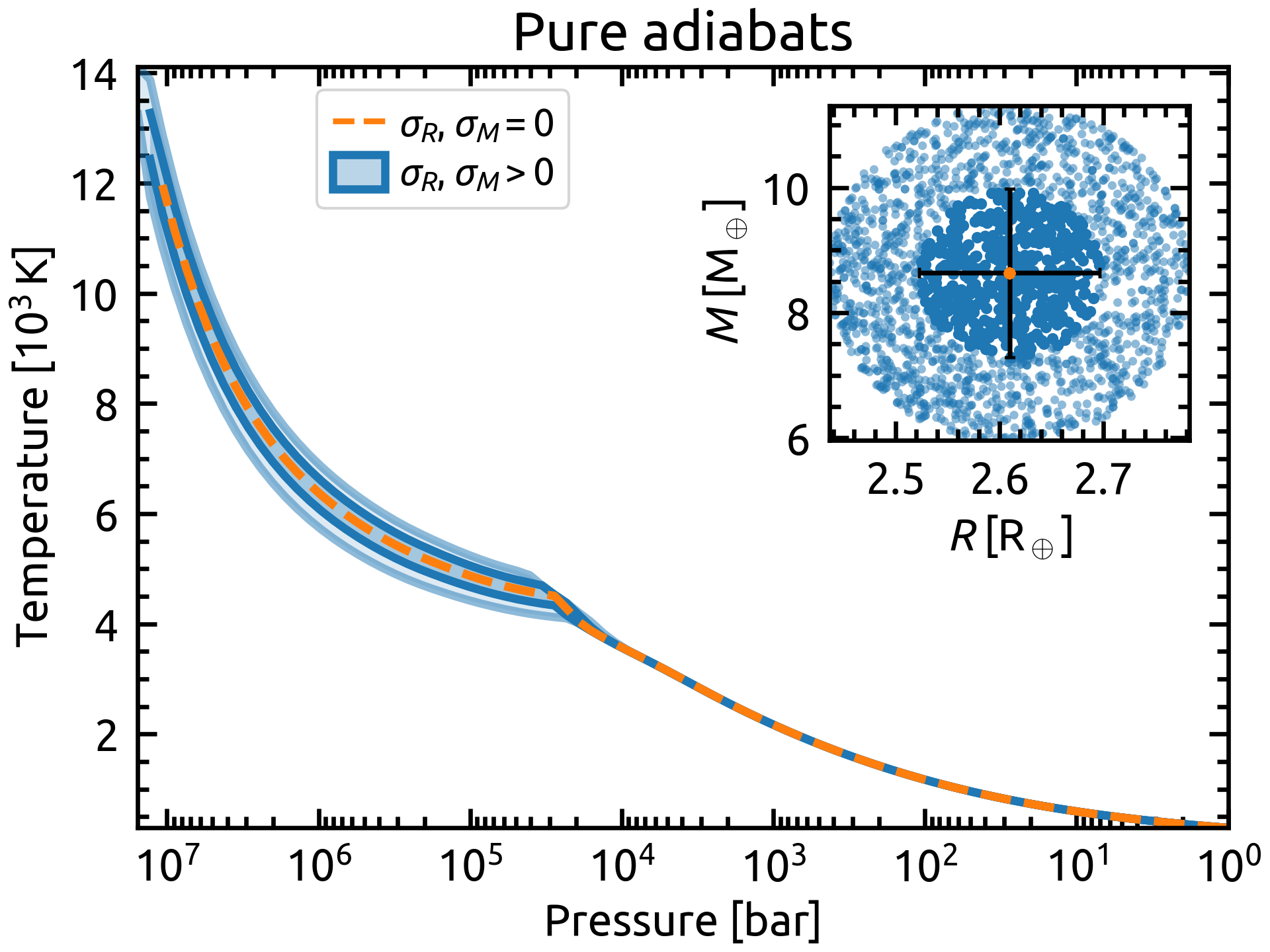}
    \includegraphics[width=0.49\linewidth]{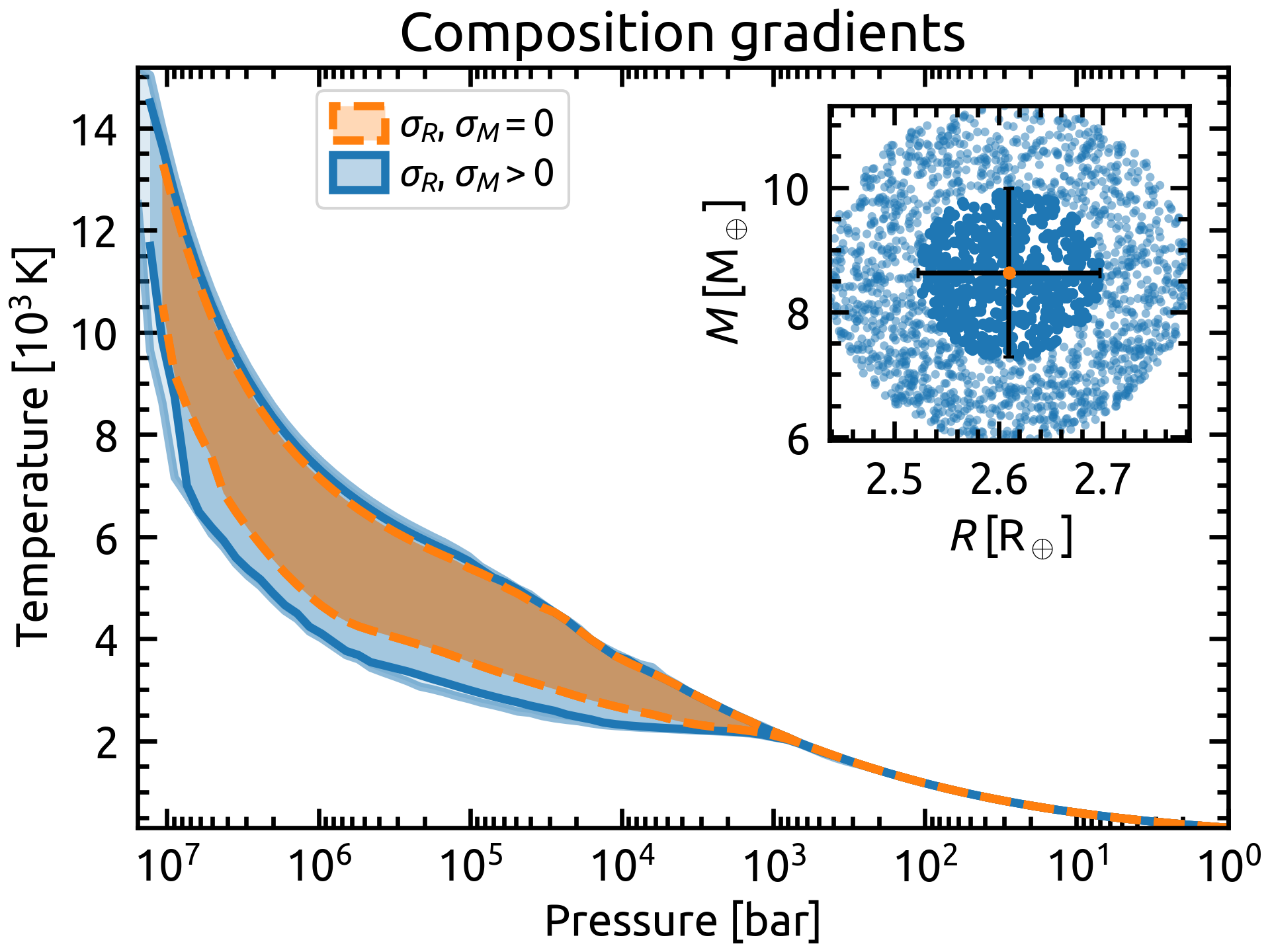}
    \caption{
    Temperature and pressure profiles for K2-18\,b models with a 1\,bar temperature of 300\,K and two materials.
    The inset panels indicate their total mass and radius, with the black error bars depicting the $\sigma_\mathrm{observation}$ from Table~\ref{tab:planet_data}.
    Left: Purely adiabatic models. Right: Composition-gradient models. 
    We display $10^4$ models that fit the measured mass and radius uncertainties $\left( \sigma_{R},\sigma_{M}>0 \right)$ as well as $10^3$ models that assume negligible measurement uncertainties $\left( \sigma_{R},\sigma_{M}=0 \right)$ in both cases.
    Since in the case of composition gradients the solutions are already degenerate for $\sigma_{R},\sigma_{M}=0$, the impact of different measurement uncertainties is limited.  
    The models within $1\sigma_\mathrm{observation}$ are not a separate set of runs but the subset of the $\sigma_R,\sigma_M>0$ models (sampled within $2\sigma_\mathrm{observation}$), whose mass and radius lie within $1\sigma_\mathrm{observation}$.
    }
    \label{fig:K2-18b_HHe-rock_TP_pure+grad-s0+s1-T300}
\end{figure*}

%%%%%%%%%%%%%%%%%%%%%%%%%%%%%%%%%%%%%%%%%%%%%%%%%%%%%%%%%%%%%%

Figure~\ref{fig:K2-18b_HHe-rock_slice_pure+grad-s0-T300} presents interior models for K2-18\,b with a radius of $R=2.61\,$R$_\oplus$ and a mass of $M=8.63\,$M$_\oplus$. 
There is one single solution for purely adiabatic models, since the assumptions of only two materials and the fixed values of $R$ and $M$ uniquely determine the interior structure and bulk composition in this case.
This is further illustrated in the left panel of Fig.~\ref{fig:K2-18b_HHe-rock_TP_pure+grad-s0+s1-T300}.  
All $10^3$ purely adiabatic solutions overlap and therefore share identical pressure ($P$) and temperature ($T$) profiles.
No degeneracy is present in this case.
The degeneracy appears only when the radius and mass are allowed to vary within their observational uncertainties. 
Then, various purely adiabatic solutions become consistent with the data and occupy distinct regions of the $P$--$T$ space. 

%%%%%%%%%%%%%%%%%%%%%%%%%%%%%%%%%%%%%%%%%%%%%%%%%%%%%%%%%%%%%%

In contrast, Fig.~\ref{fig:K2-18b_HHe-rock_slice_pure+grad-s0-T300} shows that composition-gradient models remain degenerate even when $R$ and $M$ are fixed. 
Multiple interior structures reproduce exactly the same values of $R=2.61\,$R$_\oplus$ and $M=8.63\,$M$_\oplus$. 
This is because a two-material system does not uniquely determine the location of the composition gradients, and different gradients can produce identical bulk properties if the gradient steepness is adjusted.
In addition, under this configuration H--He can mix with rock \citep[see also][]{Gilmore2026}. 
As a result, layers containing multiple materials become possible. 
The existence of mixed-composition regions introduces additional degrees of freedom and therefore further increases the number of possible solutions.

%%%%%%%%%%%%%%%%%%%%%%%%%%%%%%%%%%%%%%%%%%%%%%%%%%%%%%%%%%%%%%

The right panel of Fig.~\ref{fig:K2-18b_HHe-rock_TP_pure+grad-s0+s1-T300} clearly shows the consequences. 
The distribution of solutions obtained for non-zero observational uncertainties ($\sigma_R,\sigma_M>0$, blue) is similar to that obtained for fixed radius and mass ($\sigma_R,\sigma_M=0$, orange). 
This is because models with composition gradients are already strongly degenerate even in the absence of measurement uncertainties. 
Therefore, the dominant source of degeneracy is not observational uncertainty but the structure of the inverse problem itself \citep[in agreement with][]{Otegi2020b}. 
This conclusion is consistent with studies of the Solar System `ice giants'. 
Despite observational constraints that are orders of magnitude more precise and more diverse than those available for exoplanets, the interior structures of Uranus and Neptune remain unknown.
In particular, the relative contributions of rocky material and volatile-rich material (such as water) are still under debate \citep[e.g.][]{Neuenschwander2024, Morf2024, Morf2025, Ramirez2026}.

%%%%%%%%%%%%%%%%%%%%%%%%%%%%%%%%%%%%%%%%%%%%%%%%%%%%%%%%%%%%%%

\subsection{Inferred composition using four materials}
\label{sec:Results_4_materials}

%%%%%%%%%%%%%%%%%%%%%%%%%%%%%%%%%%%%%%%%%%%%%%%%%%%%%%%%%%%%%%

\begin{figure*}
    \centering
    \includegraphics[width=0.49\linewidth]{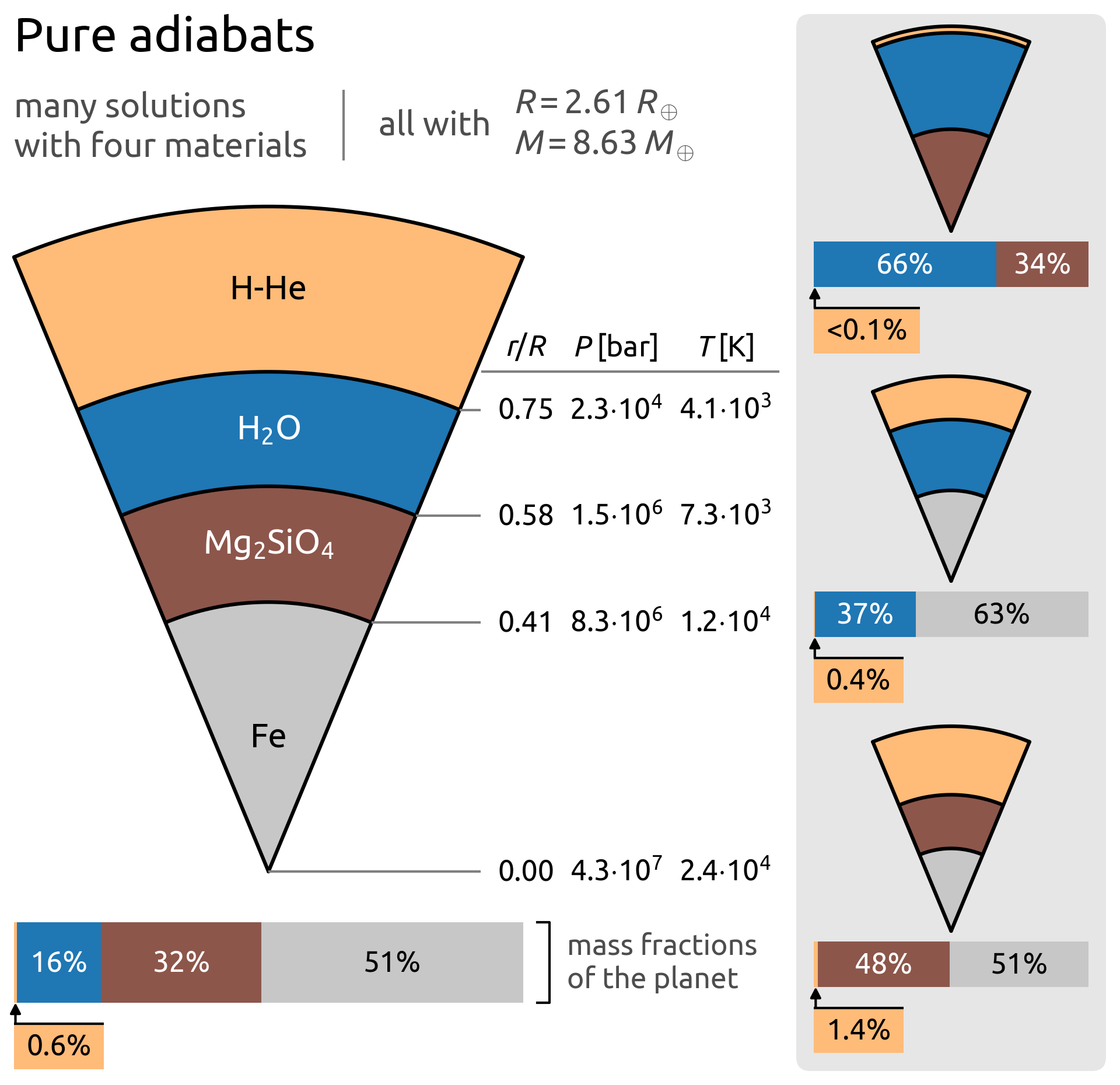}
    \includegraphics[width=0.49\linewidth]{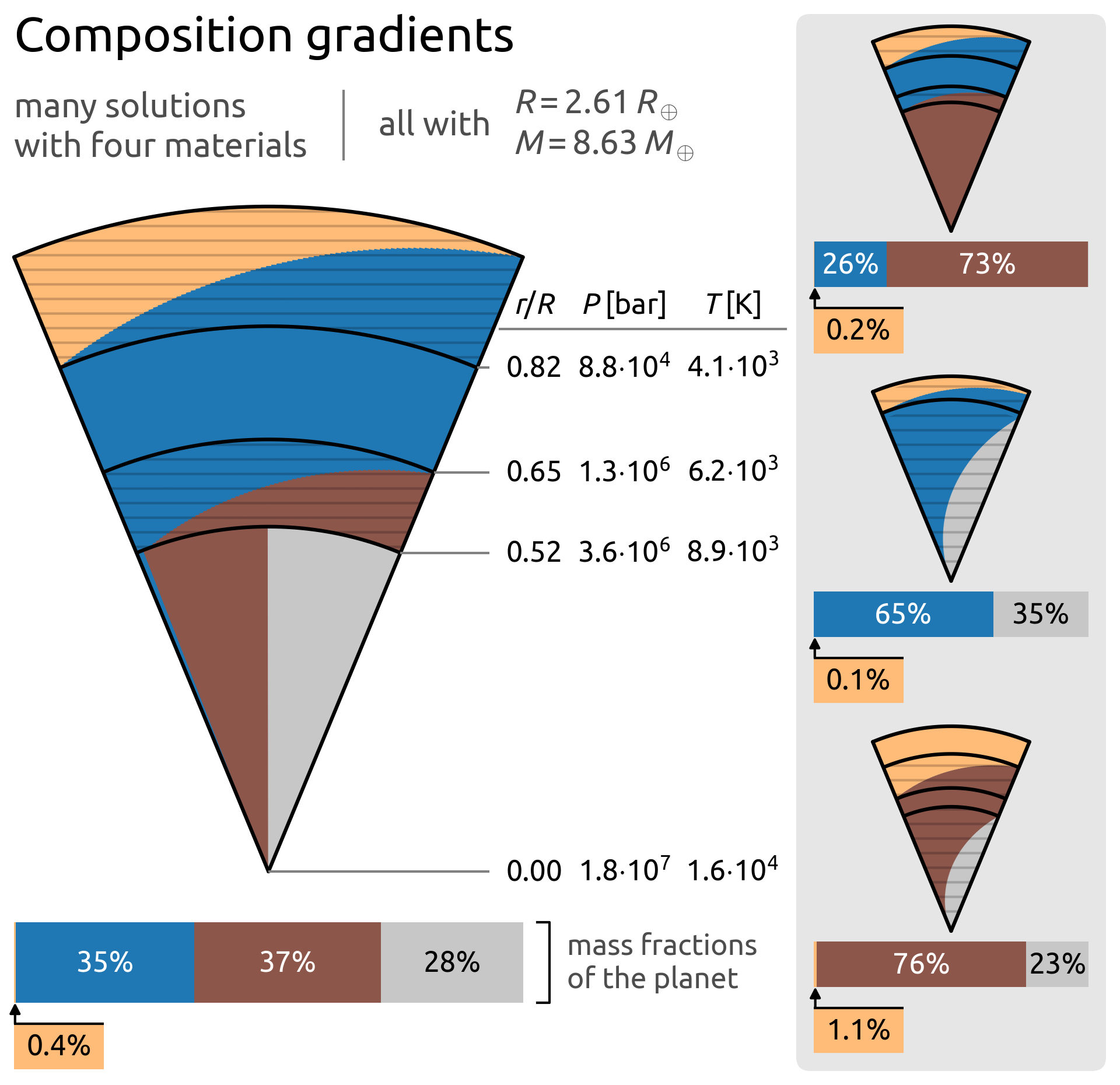}
    \caption{
    Same as Fig.~\ref{fig:K2-18b_HHe-rock_slice_pure+grad-s0-T300} but for interior models that assume four materials. 
    All the solutions have the same radius $R$ and mass $M$. 
    Both purely adiabatic and composition-gradient models allow for a wide range of interior solutions due to higher degeneracy. 
    }
    \label{fig:K2-18b_slice_pure+grad-s0-T300}
\end{figure*}

\begin{figure*}
    \centering
    \includegraphics[width=0.49\linewidth]{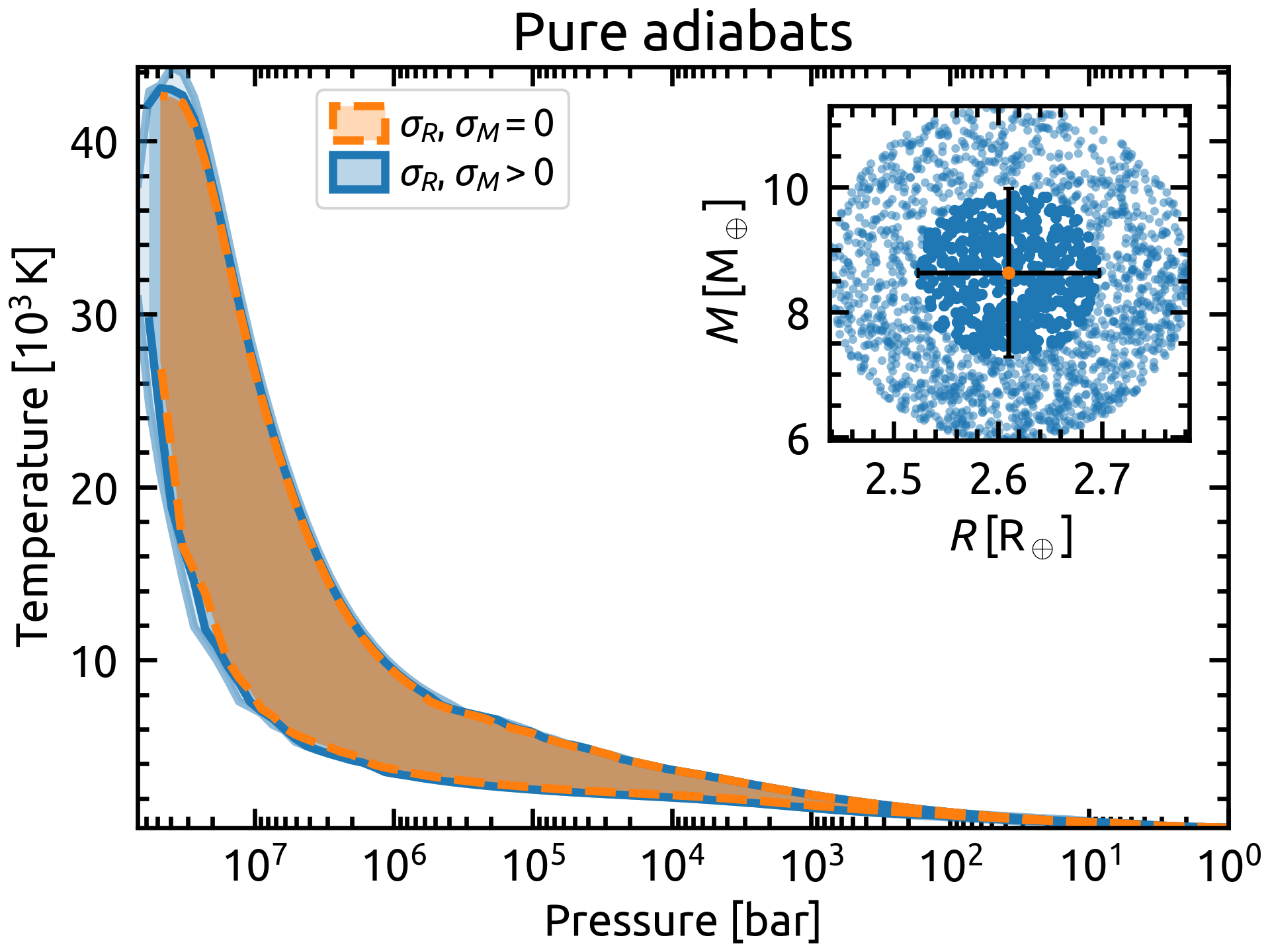}
    \includegraphics[width=0.49\linewidth]{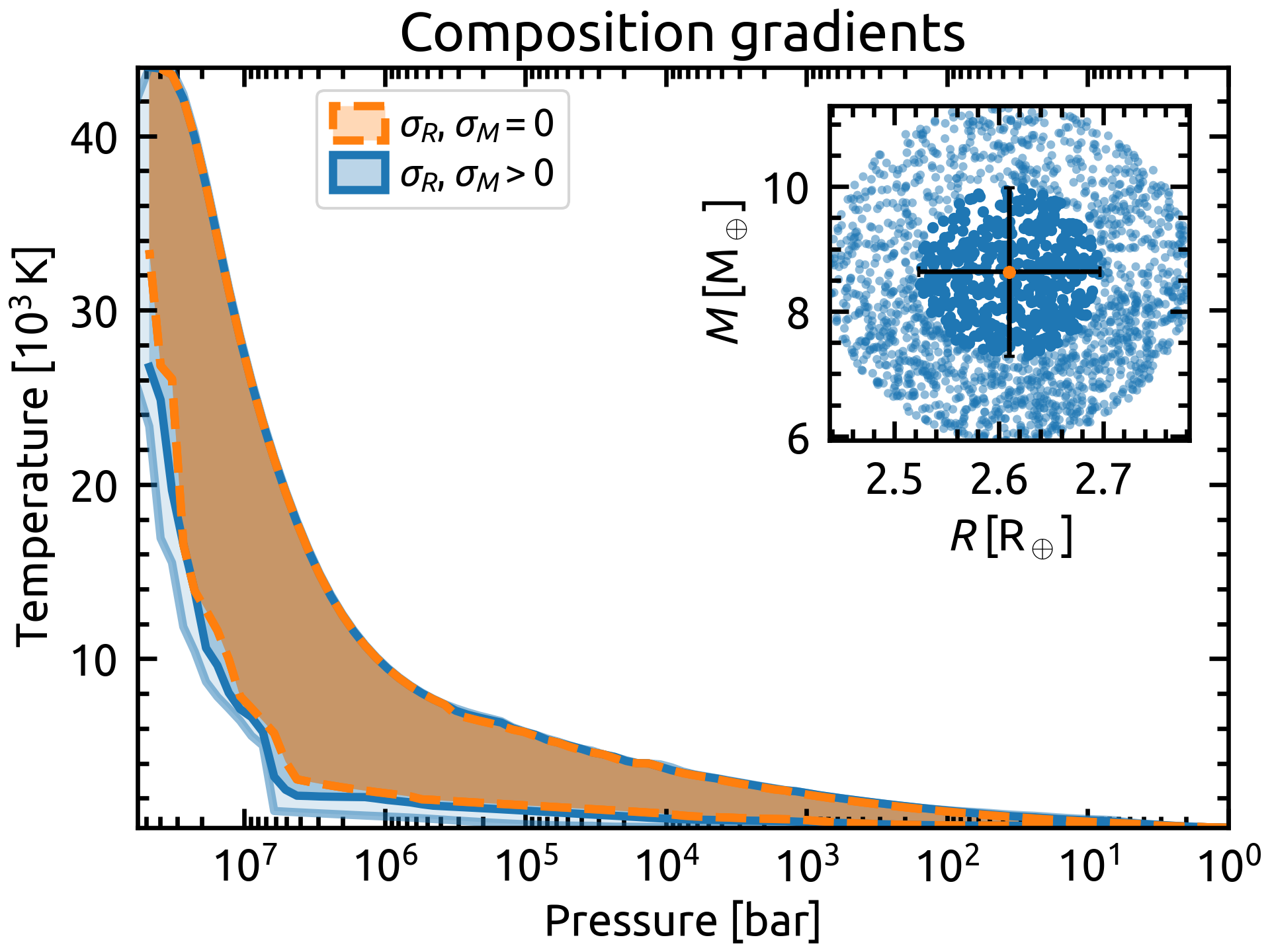}
    \caption{
        Same as Fig.~\ref{fig:K2-18b_HHe-rock_TP_pure+grad-s0+s1-T300} but assuming four materials.
        Models with negligible measurement uncertainties ($\sigma_R,\sigma_M=0$) already span a similar pressure--temperature range as models within $1\sigma_\mathrm{observation}$ and $2\sigma_\mathrm{observation}$ for both purely adiabatic and composition-gradient models.
    }
    \label{fig:K2-18b_TP_pure+grad-s0+s1-T300}
\end{figure*}

%%%%%%%%%%%%%%%%%%%%%%%%%%%%%%%%%%%%%%%%%%%%%%%%%%%%%%%%%%%%%%

Figure~\ref{fig:K2-18b_slice_pure+grad-s0-T300} shows the same analysis as Fig.~\ref{fig:K2-18b_HHe-rock_slice_pure+grad-s0-T300} but assumes four materials in the planetary interior models. 
In addition to H--He and rock, water and iron are included as possible constituents. 
While real planets are expected to contain an even wider range of chemical species, these four materials span a broad range of densities relevant for interior modelling and hence are a compromise between being overly simplistic and considering too many degrees of freedom. 

%%%%%%%%%%%%%%%%%%%%%%%%%%%%%%%%%%%%%%%%%%%%%%%%%%%%%%%%%%%%%%

The results for the inferred composition and structure exhibit a behaviour that differs substantially from the two-material case. 
For fixed values of $R=2.61\,$R$_\oplus$ and $M=8.63\,$M$_\oplus$, numerous solutions become possible even for the purely adiabatic models. 
Consequently, the strong degeneracy is no longer restricted to composition-gradient models.
Furthermore, models with vanishing observational uncertainties now span a similar range in the $P$--$T$ space as models within $2\sigma_\mathrm{observation}$ for both purely adiabatic and composition-gradient models, as shown in Fig.~\ref{fig:K2-18b_TP_pure+grad-s0+s1-T300}.
Therefore, both purely adiabatic models and composition-gradient models clearly show that improving the uncertainties of the mass and radius measurements does not necessarily further constrain the planetary interior.

%%%%%%%%%%%%%%%%%%%%%%%%%%%%%%%%%%%%%%%%%%%%%%%%%%%%%%%%%%%%%%

Table~\ref{tab:Abundances_K2-18b_300K} lists the inferred minimum and maximum values of the (combined) rock and iron mass fractions for different assumed observational uncertainties.
When $\sigma_R,\sigma_M>0$ is considered, a wider range of possibilities, such as solutions with lower rock and iron mass fractions or higher H--He mass fractions, is permitted.
However, regardless of the observational uncertainties, models with composition gradients always cover more possibilities compared to purely adiabatic models.
For example, in the $\sigma_R,\sigma_M=0$ case, composition-gradient models can have rock and iron mass fractions of 0\%, while purely adiabatic models never go below 10\%.

%%%%%%%%%%%%%%%%%%%%%%%%%%%%%%%%%%%%%%%%%%%%%%%%%%%%%%%%%%%%%%

For purely adiabatic models, the maximum H--He mass fraction remains at or below 3\%, independent of the assumed measurement uncertainties.
Interior models with composition gradients permit substantially larger values. 
The maximum H--He mass fraction can reach 9\%, corresponding to an increase by a factor of three relative to the purely adiabatic solutions. 
At 600\,K (for K2-18\,b), the maximum H--He mass fraction reaches 10\% for composition-gradient models compared with 1\% for purely adiabatic models, corresponding to a factor of $\sim$10.
Solutions with composition gradients therefore again populate regions of the parameter space that are inaccessible to purely adiabatic models. 
They provide additional freedom in the redistribution of materials throughout the interior and consequently allow for a wider range of bulk compositions. 
On the other hand, the sample of purely adiabatic models corresponds to a restricted subset of the possible solution space. 
As a result, conclusions derived exclusively from purely adiabatic models are likely to underestimate the true range of possible interior structures and bulk compositions.  
In particular, they may lead to overly restrictive statements regarding the composition and formation history of sub-Neptunes. 

%%%%%%%%%%%%%%%%%%%%%%%%%%%%%%%%%%%%%%%%%%%%%%%%%%%%%%%%%%%%%%

Finally, Table~\ref{tab:Abundances_K2-18b_300K} lists the results from \cite{Madhusudhan2020, Howard2025, Schmidt2025} for comparison. 
The solutions presented by previous studies mostly correspond to the purely adiabatic case with $\sigma_R,\sigma_M>0$. 
Differences in the maximum H--He mass fractions can be attributed to different thermal assumptions. 
For example, variations in the assumed 1\,bar temperature or the inclusion of extended isothermal layers in the outer envelope can significantly affect the inferred mass of low-density material and therefore increase the total H--He abundance. 

%%%%%%%%%%%%%%%%%%%%%%%%%%%%%%%%%%%%%%%%%%%%%%%%%%%%%%%%%%%%%%

\begin{table}
\caption{
Total H--He as well as rock and iron mass fractions for K2-18\,b models with four materials.
}
\centering
\setlength{\tabcolsep}{5.5pt}
    \begin{tabular}{lllcccc}
    \hline
    \hline
    K2-18\,b & $\sigma_{R},\sigma_{M}$ & $T_{1\,\mathrm{bar}}$ & \multicolumn{2}{c}{\tiny{H--He [\%]}} & \multicolumn{2}{c}{\tiny{Mg$_2$SiO$_4$+Fe [\%]}} \\
    \cline{4-5} \cline{6-7}
    & & & \tiny{min} & \tiny{max} & \tiny{min} & \tiny{max} \\
    \hline \hline
    \multirow{2}{*}{\makecell[l]{Pure \\ adiabats}} & =0 & 300\,K & 0 & 3  & 10  & 99 \\
    & >0 & 300\,K & 0 & 3  & 0 & 100 \\
    \hline
    \multirow{2}{*}{\makecell[l]{Composition \\ gradients}} & =0 & 300\,K & 0 & 4 & 0 & 100 \\
    & >0 & 300\,K & 0 & 9 & 0 & 100 \\
    \hline \hline
    Mad+20 & >0 & & 0 & 5 & 0 & 99 \\
    \hline
    Sch+25 & >0 & & 0 & 6 & 0 & 95 \\
    \hline
    How+25 & >0 & & 0 & 6 & 0 & 94 \\
    \hline \hline
    \end{tabular}
\label{tab:Abundances_K2-18b_300K}
\tablefoot{
The findings by \cite{Madhusudhan2020, Howard2025, Schmidt2025} are included as a comparison.
}
\end{table}

%%%%%%%%%%%%%%%%%%%%%%%%%%%%%%%%%%%%%%%%%%%%%%%%%%%%%%%%%%%%%%

\subsection{Importance of atmosphere models and stellar constraints}
\label{sec:Results_atmosphere_hoststar}
 
%%%%%%%%%%%%%%%%%%%%%%%%%%%%%%%%%%%%%%%%%%%%%%%%%%%%%%%%%%%%%%

We next consider two additional sources of information that are frequently used to constrain exoplanet interior models: the atmospheric properties and host-star abundances. 
Both are commonly invoked to reduce the degeneracy inherent to exoplanet interior inference. 

%%%%%%%%%%%%%%%%%%%%%%%%%%%%%%%%%%%%%%%%%%%%%%%%%%%%%%%%%%%%%%

\begin{figure}
    \centering
    \includegraphics[width=\linewidth]{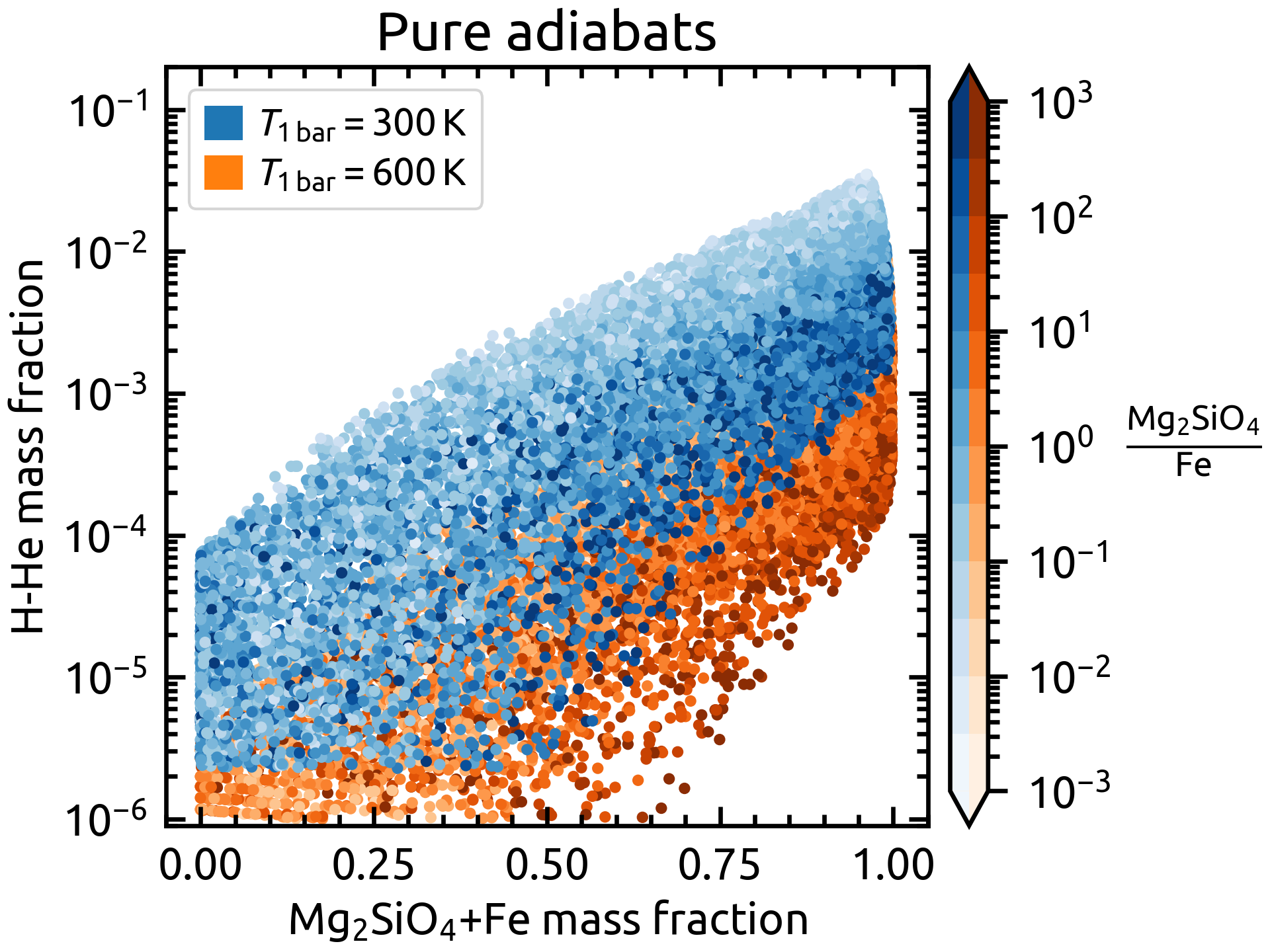}
    \caption{
    Total H--He as well as rock and iron mass fractions for purely adiabatic K2-18\,b models that fit the measured mass and radius uncertainties ($\sigma_{R},\sigma_{M}>0$).  
    We include $10^4$ models with a 1\,bar temperature of 300\,K and $10^4$ models with a 1\,bar temperature of 600\,K.
    The dark and light points correspond to rock- and iron-dominated solutions, respectively. 
    }
    \label{fig:K2-18b_hhe-vs-rock_pure-s1-T300+T600_by-Mg2SiO4-Fe}

    \vspace{0.5cm}
    
    \centering
    \includegraphics[width=\linewidth]{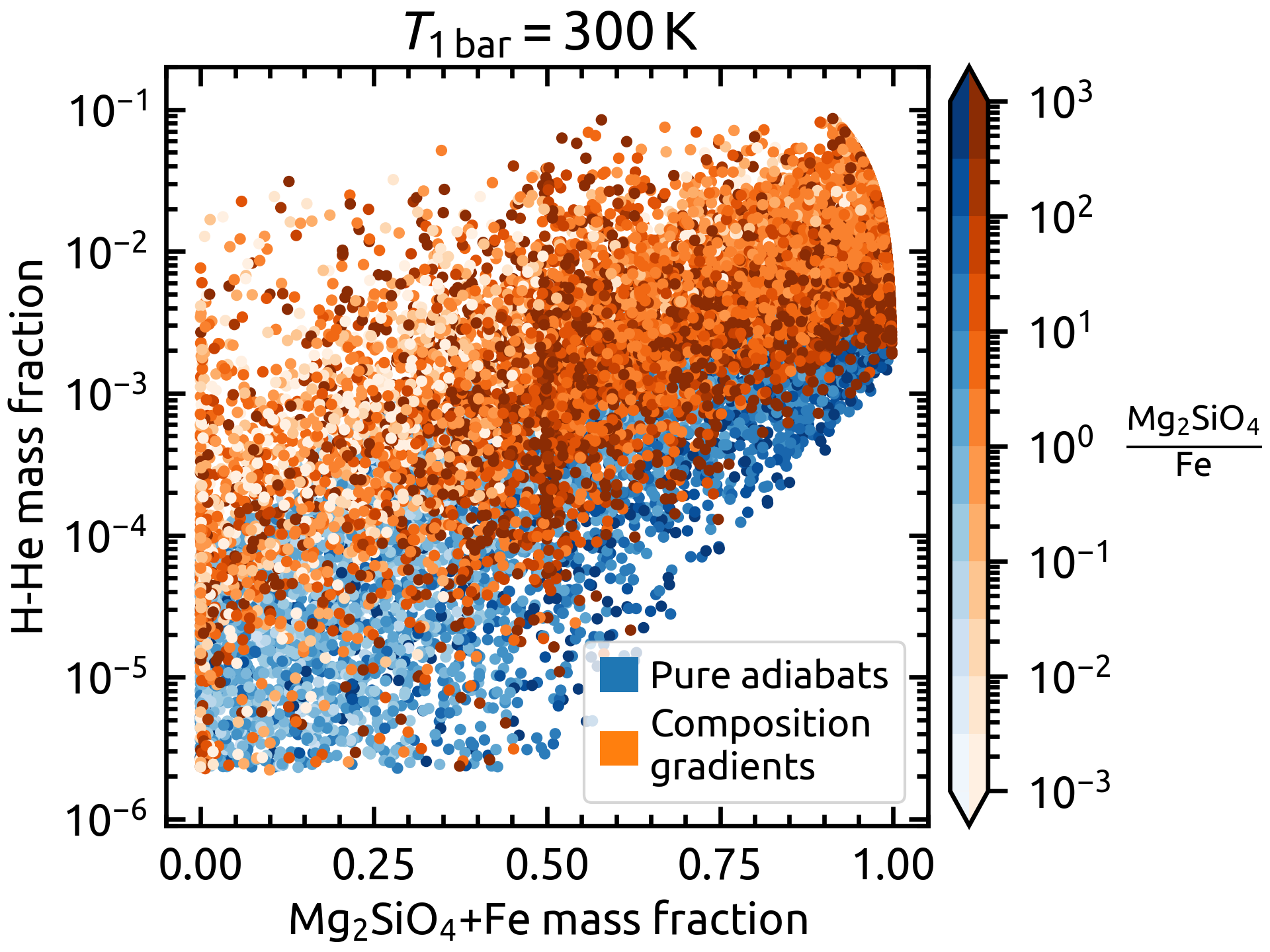}
    \caption{
    Same as Fig.~\ref{fig:K2-18b_hhe-vs-rock_pure-s1-T300+T600_by-Mg2SiO4-Fe} but showing two datasets with a 1\,bar temperature of 300\,K. 
    The blue dataset is as presented in Fig.~\ref{fig:K2-18b_hhe-vs-rock_pure-s1-T300+T600_by-Mg2SiO4-Fe}, while the orange dataset now depicts the solutions of models with composition gradients. 
    We find that composition-gradient models show a weaker H--He vs Mg$_2$SiO$_4$/Fe correlation. 
    }
    \label{fig:K2-18b_hhe-vs-rock_pure+grad-s1-T300_by-Mg2SiO4-Fe}
\end{figure}

Figure~\ref{fig:K2-18b_hhe-vs-rock_pure-s1-T300+T600_by-Mg2SiO4-Fe} shows solutions for purely adiabatic models of K2-18\,b assuming two different atmospheric boundary conditions. 
The blue points correspond to a 1\,bar temperature of 300\,K, while the orange points correspond to a 1\,bar temperature of 600\,K. 
In both cases, the radius and mass are allowed to vary within their observational uncertainties ($\sigma_R,\sigma_M>0$). 
We observe a clear shift between the two populations: 
models with higher 1\,bar temperatures require lower H--He mass fractions. 
This is because higher temperatures increase the specific volume of the envelope material, and as a result, less H--He is required to reproduce the observed planetary radius. 
Conversely, colder atmospheres require larger amounts of low-density material. 
The inferred bulk composition is thus sensitive to the assumed atmospheric temperature structure \citep[see also][]{Otegi2020b}. 
Our analysis demonstrates that atmospheric constraints can eliminate specific regions of the interior parameter space for purely adiabatic models.

%%%%%%%%%%%%%%%%%%%%%%%%%%%%%%%%%%%%%%%%%%%%%%%%%%%%%%%%%%%%%%

%%%%%%%%%%%%%%%%%%%%%%%%%%%%%%%%%%%%%%%%%%%%%%%%%%%%%%%%%%%%%%

Figure~\ref{fig:K2-18b_hhe-vs-rock_pure-s1-T300+T600_by-Mg2SiO4-Fe} also reveals a relationship between the H--He abundance and the rock-to-iron (Mg$_2$SiO$_4$/Fe) ratio, which is indicated by the colour scale. 
A clear trend is visible: models with larger H--He mass fractions exhibit smaller rock-to-iron ratios. 
The Spearman rank correlation coefficients for H--He versus Mg$_2$SiO$_4$/Fe are found to be $-0.33$ for the 300\,K models and $-0.36$ for the 600\,K models; see also Fig.~\ref{fig:K2-18b_corr}.
If the planetary elemental abundance ratios are the same as in the host star, measurements of the analogous stellar rock-to-iron ratio would restrict the permitted range of possible interior solutions, as shown in Fig.~\ref{fig:K2-18b_hhe-vs-rock_pure-s1-T300+T600_by-Mg2SiO4-Fe}. 
In the simplified purely adiabatic framework, stellar abundances could therefore be a significant additional constraint for interior models \citep[e.g.][]{Dorn2017}.
However, it is unclear whether this assumption is justified given the available measurements \citep[e.g.][]{Plotnykov2020, Schulze2021, Brinkman2024}.
Even in the Solar System, the terrestrial planets are found to have different abundance ratios (for example Mercury versus Earth). 
The case of Neptunes and sub-Neptunes is even more uncertain.
Therefore, it remains unknown whether (and under what conditions) the elemental ratios in planets can be assumed to be similar to those of their host stars; see also \cite{Teske2024} for a review.

%%%%%%%%%%%%%%%%%%%%%%%%%%%%%%%%%%%%%%%%%%%%%%%%%%%%%%%%%%%%%%

\begin{figure*}
    \centering
    \includegraphics[width=0.49\linewidth]{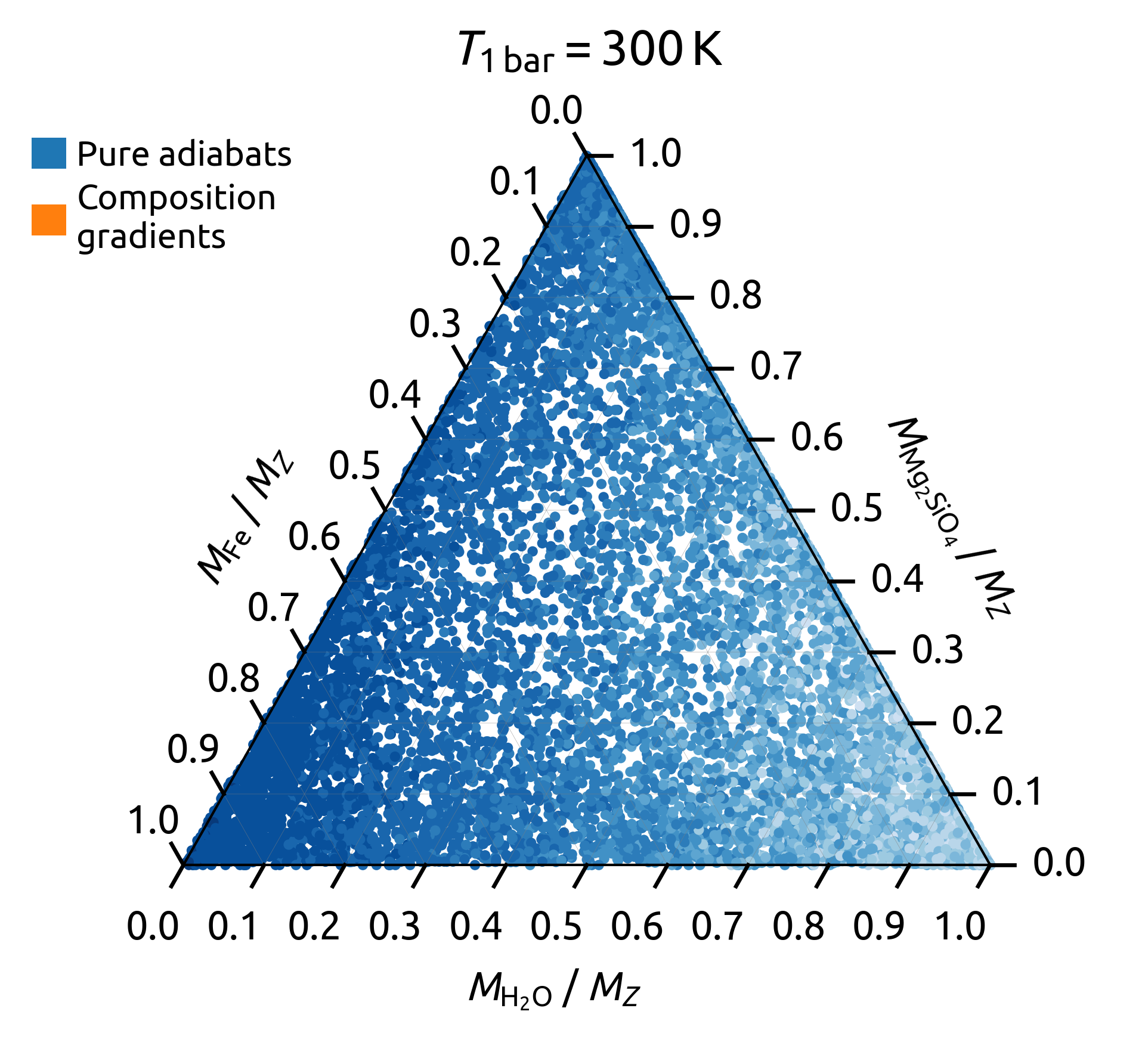}
    \includegraphics[width=0.49\linewidth]{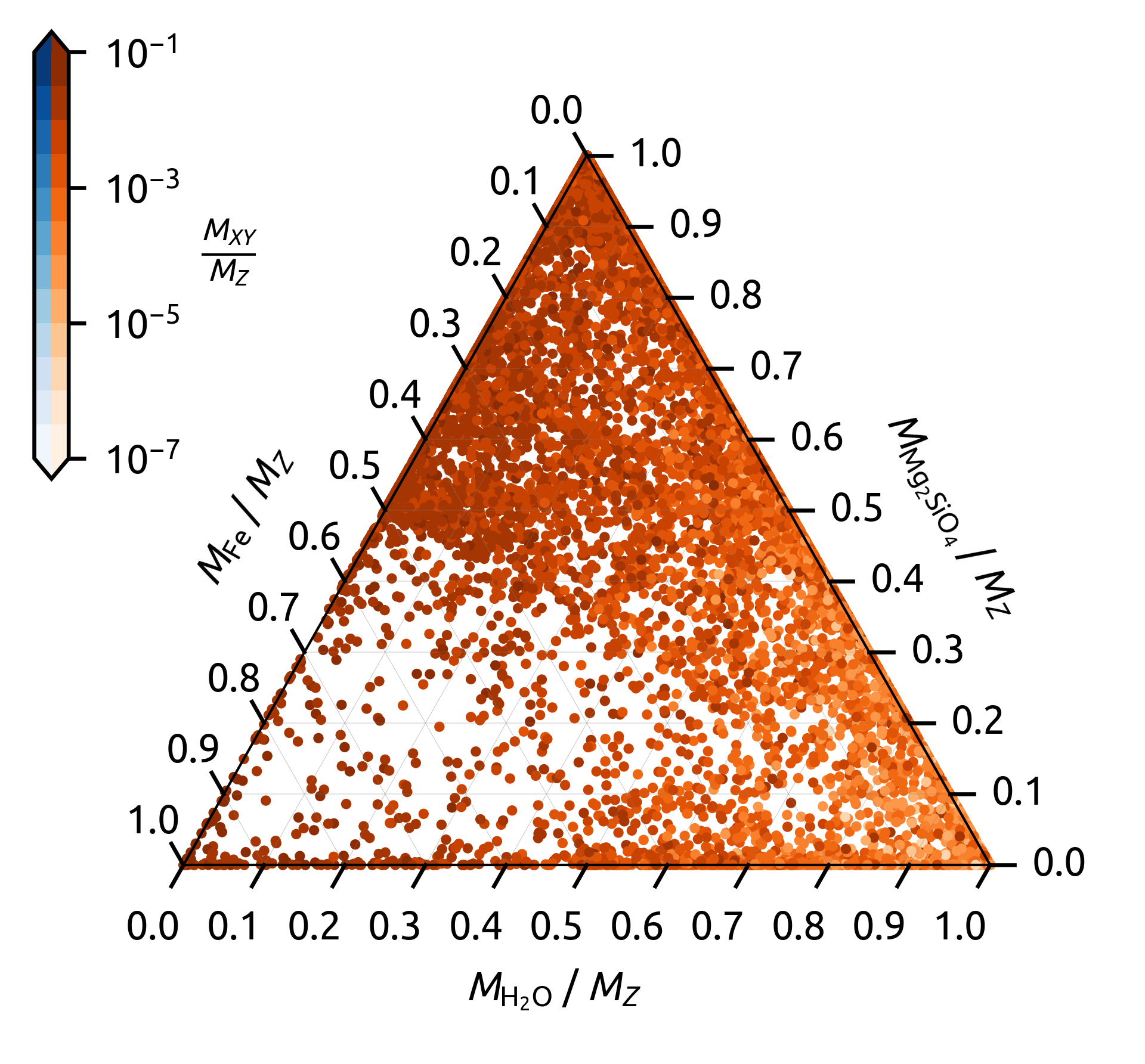}
    \caption{
    Total water (H$_2$O), rock (Mg$_2$SiO$_4$), and iron (Fe) mass fractions for K2-18\,b models with a 1\,bar temperature of 300\,K.
    The mass fractions are normalised by $M_Z=M_{\mathrm{H}_2\mathrm{O}}+M_{\mathrm{Mg}_2\mathrm{SiO}_4}+M_\mathrm{Fe}$ and hence add up to unity. 
    The colour scale shows the ratio of the total H--He mass, $M_{XY}$, to $M_Z$ on a logarithmic scale, with the dark points corresponding to larger ratios.  
    The left ternary diagram shows $10^4$ purely adiabatic models, while the right ternary diagram shows $10^4$ models with composition gradients. 
    }
    \label{fig:K2-18b_ternary_pure+grad-s1-T300}
\end{figure*}

\begin{figure}
    \centering
    \includegraphics[width=\linewidth]{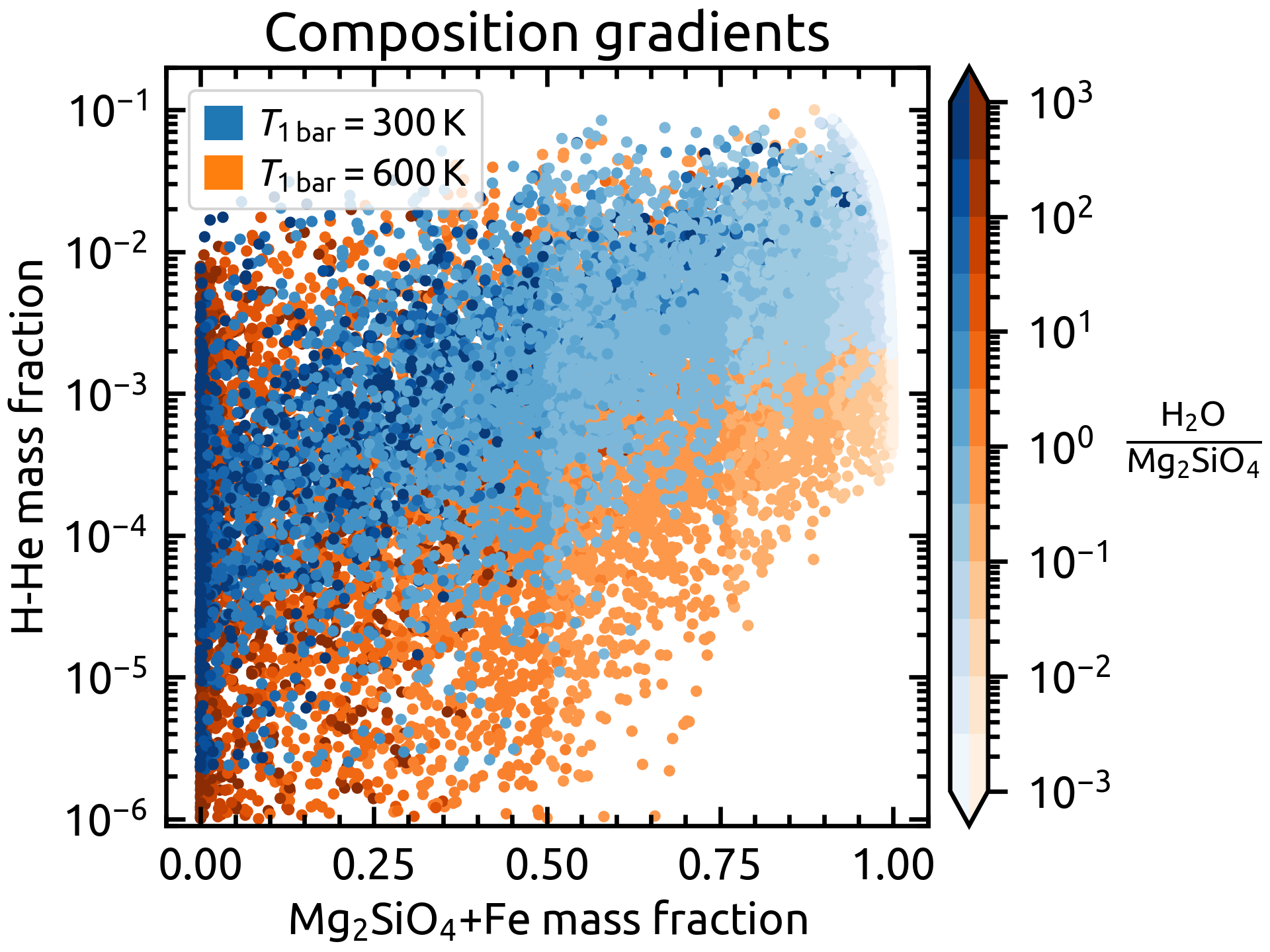}
    \caption{
    Same as Fig.~\ref{fig:K2-18b_hhe-vs-rock_pure-s1-T300+T600_by-Mg2SiO4-Fe} but considering models with composition gradients. 
    The dark and light points show water-dominated and rock-dominated solutions, respectively. 
    Models with composition gradients show a H--He vs H$_2$O/Mg$_2$SiO$_4$ correlation.
    }
    \label{fig:K2-18b_hhe-vs-rock_grad-s1-T300+T600_by-H2O-Mg2SiO4}
\end{figure}

%%%%%%%%%%%%%%%%%%%%%%%%%%%%%%%%%%%%%%%%%%%%%%%%%%%%%%%%%%%%%%

The situation changes when interior composition gradients are considered. 
Figure~\ref{fig:K2-18b_hhe-vs-rock_pure+grad-s1-T300_by-Mg2SiO4-Fe} directly compares purely adiabatic and composition-gradient models. 
First, we again find that solutions for models with composition gradients occupy a much larger region of the parameter space and that a wider range of H--He abundances is possible.  
Furthermore, the H--He versus Mg$_2$SiO$_4$/Fe correlations diminish substantially.
For models with composition gradients, the Spearman rank correlation coefficients are $-0.06$ ($T_{1\,\mathrm{bar}}=300\,$K) and $-0.07$ ($T_{1\,\mathrm{bar}}=600\,$K), as also demonstrated in Fig.~\ref{fig:K2-18b_corr}.
The physical interpretation for this reduction is rather straightforward:
composition gradients introduce additional internal degrees of freedom.
Consequently, allowing for composition gradients can break down the correlations seen in purely adiabatic models, making certain quantities more independent.
Therefore, additional constraints from stellar abundances provide less constraining power.

%%%%%%%%%%%%%%%%%%%%%%%%%%%%%%%%%%%%%%%%%%%%%%%%%%%%%%%%%%%%%%

\subsection{Correlations between different elemental abundances}
\label{sec:Results_ice_rock_correlations}

%%%%%%%%%%%%%%%%%%%%%%%%%%%%%%%%%%%%%%%%%%%%%%%%%%%%%%%%%%%%%%

Next, we investigate the inferred ice-to-rock ratios and further correlations, such as H--He versus Fe. 
Figure~\ref{fig:K2-18b_ternary_pure+grad-s1-T300} compares purely adiabatic and composition-gradient models for K2-18\,b. 
We consider the low-temperature scenario ($T_{1\,\mathrm{bar}}=300\,$K) and include all the solutions that are consistent with the observational uncertainties of the mass and radius ($\sigma_R,\sigma_M>0$ case). 
The ternary diagrams display the inferred relative abundances of water, rock, and iron. 
We find that in both model classes, all pure end-members (planets made of pure water, pure rock, or pure iron) are possible.
For composition-gradient models, our methods generate fewer iron-dominated solutions. 
This feature reflects our sampling procedure, and therefore it should not be concluded that such models are less likely. 

%%%%%%%%%%%%%%%%%%%%%%%%%%%%%%%%%%%%%%%%%%%%%%%%%%%%%%%%%%%%%%

\begin{figure}
    \centering
    \includegraphics[width=\linewidth]{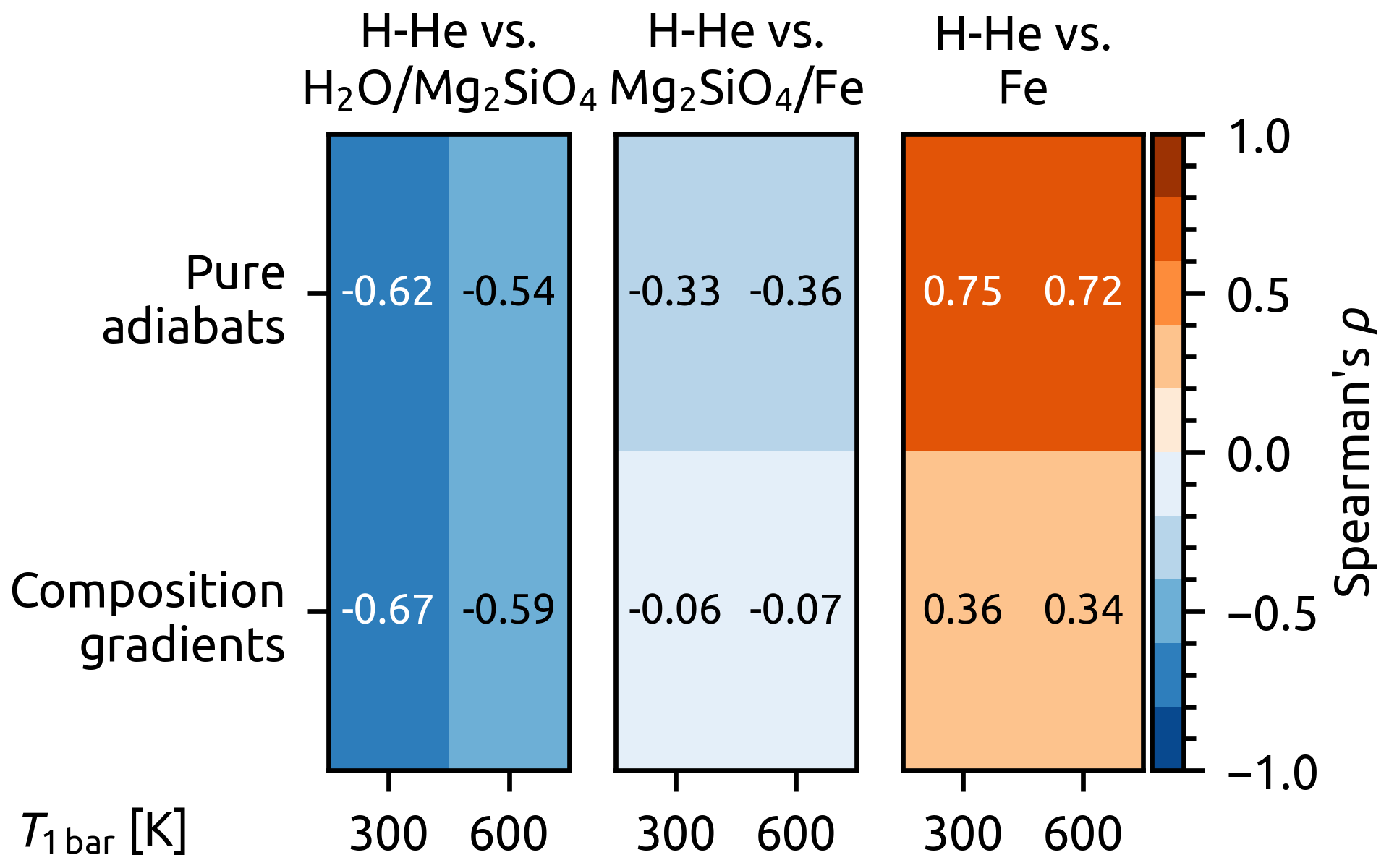}
    \caption{
    Spearman rank correlation coefficients for K2-18\,b models consistent with the measurement uncertainties ($\sigma_R,\sigma_M>0$ case). 
    The results for H--He vs H$_2$O/Mg$_2$SiO$_4$, H--He vs Mg$_2$SiO$_4$/Fe, and H--He vs Fe correlations are shown for $10^4$ purely adiabatic and $10^4$ composition-gradient models and each 1\,bar temperature case.
    The inferred $p$ values, which are smaller than $10^{-8}$, are valid only within our modelling framework (Sect.~\ref{sec:Limitations}).
    Models that include composition gradients can reduce the correlations seen in purely adiabatic models.
    }
    \label{fig:K2-18b_corr}
\end{figure}

The colour scale in Fig.~\ref{fig:K2-18b_ternary_pure+grad-s1-T300} represents the H--He mass relative to the total heavy-element mass, $M_{XY}/M_Z$ (on a logarithmic scale), with darker colours corresponding to larger values.
For purely adiabatic models, higher H--He abundances occur almost exclusively for iron-rich interiors.
Therefore, a strong positive correlation is present between the H--He mass fraction and the iron content, quantified by a Spearman rank correlation coefficient of $0.75$. 
Figure~\ref{fig:K2-18b_corr} summarises all correlation coefficients discussed in this subsection.

%%%%%%%%%%%%%%%%%%%%%%%%%%%%%%%%%%%%%%%%%%%%%%%%%%%%%%%%%%%%%%

Interior models with composition gradients exhibit a different behaviour. 
High H--He mass fractions are found for a much broader range of iron abundances. 
The dependence on the iron content is substantially weaker, reflected by a lower Spearman rank correlation coefficient of 0.36. 
Models with composition gradients hence again somewhat decouple quantities that appear strongly connected in purely adiabatic models and reveal that, in reality, a larger set of admissible interior structures is possible. 

%%%%%%%%%%%%%%%%%%%%%%%%%%%%%%%%%%%%%%%%%%%%%%%%%%%%%%%%%%%%%%

%%%%%%%%%%%%%%%%%%%%%%%%%%%%%%%%%%%%%%%%%%%%%%%%%%%%%%%%%%%%%%

Another quantity of interest is the ice-to-rock (H$_2$O/Mg$_2$SiO$_4$) ratio. 
This property is important for understanding the interiors of Uranus and Neptune, where the relative contributions of volatile-rich and rocky material remain uncertain \citep[e.g.][]{Neuenschwander2024, Morf2024, Morf2025, Ramirez2026}. 
The inferred ice-to-rock ratio is presented in Fig.~\ref{fig:K2-18b_hhe-vs-rock_grad-s1-T300+T600_by-H2O-Mg2SiO4} for composition-gradient models. 
The Spearman rank correlation coefficients for H--He versus H$_2$O/Mg$_2$SiO$_4$ are found to be $-0.67$ ($T_{1\,\mathrm{bar}}=300\,$K) and $-0.59$ ($T_{1\,\mathrm{bar}}=600\,$K), respectively, for interior models with composition gradients. 
Both values indicate a clear negative correlation: solutions with larger H--He abundances tend to exhibit smaller ice-to-rock ratios.
Interestingly, the correlation is present in both the purely adiabatic and composition-gradient models, as can be seen in Fig.~\ref{fig:K2-18b_corr}. 
This demonstrates that composition gradients allow certain relationships to remain significant, although they encompass a substantially broader region of the solution space.
Independent constraints on the ice-to-rock ratio (from formation and evolution models) can therefore help to reduce some of the interior degeneracy in both purely adiabatic and composition-gradient models. 

%%%%%%%%%%%%%%%%%%%%%%%%%%%%%%%%%%%%%%%%%%%%%%%%%%%%%%%%%%%%%%

\section{Limitations}
\label{sec:Limitations}

This study applies two self-consistent frameworks (purely adiabatic and composition-gradient models) to planetary interiors of two sub-Neptunes. 
Several limitations remain and should be considered when interpreting the results.

%%%%%%%%%%%%%%%%%%%%%%%%%%%%%%%%%%%%%%%%%%%%%%%%%%%%%%%%%%%%%%

\begin{figure}
    \centering
    \includegraphics[width=\linewidth]{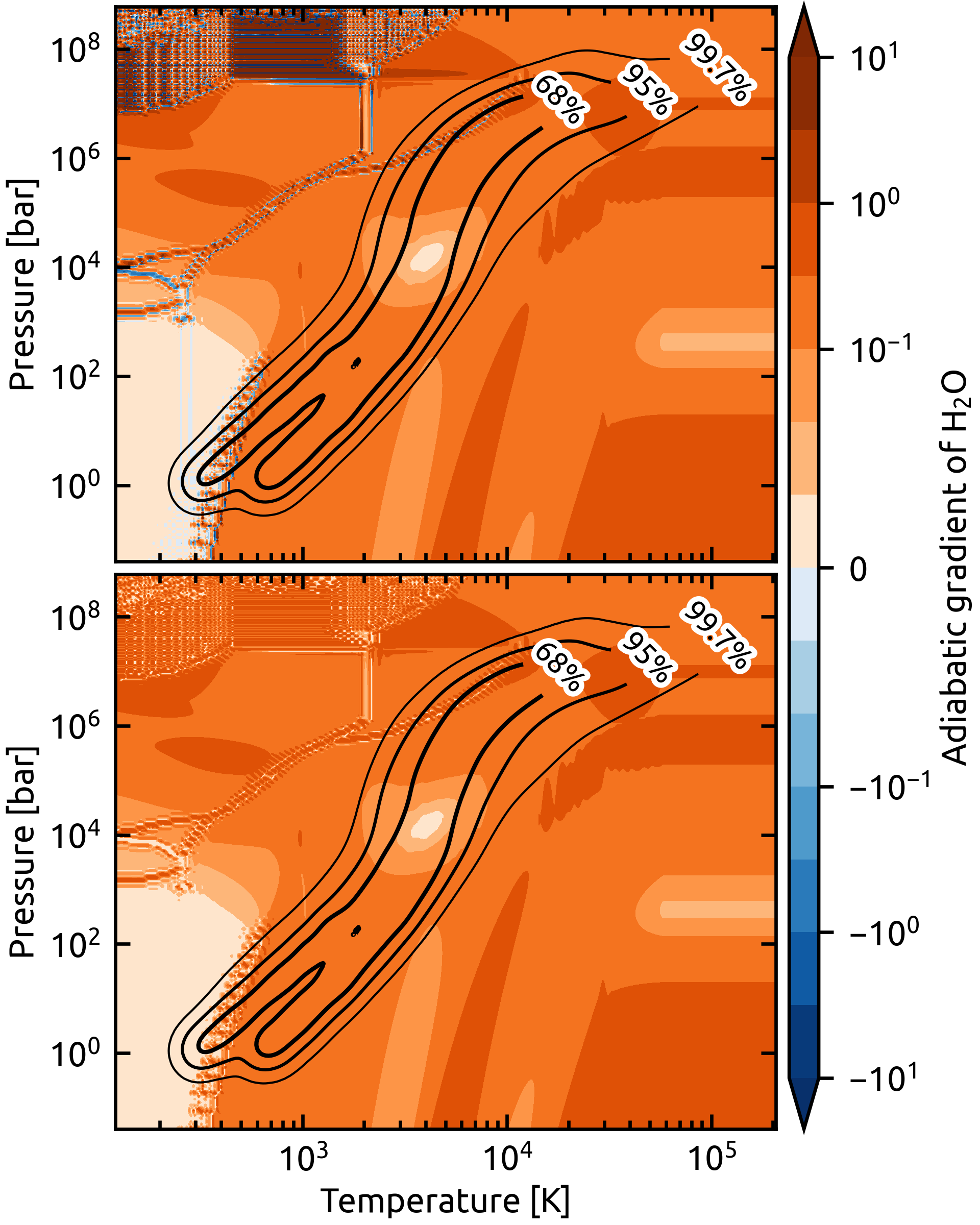}
    \caption{
    Adiabatic temperature gradient of pure water as a function of temperature and pressure.
    Top: Gradient computed directly from the EoS \citep{CanoAmoros2026}. Bottom: Same quantity following the clipping procedure described in Eq.~\eqref{eq:clipping}.
    The black contours enclose the smallest regions containing 68\%, 95\%, and 99.7\% of the pressure--temperature profiles of all computed interior models (K2-18\,b and TOI-270\,d).
    }
    \label{fig:eos-grad-ad_H2O}
\end{figure}

First, uncertainties in the equation of state (EoS) and the assumption of ideal mixing affect the inferred compositions.
The employed EoSs at planetary conditions \citep{Stewart2020, Chabrier2021, CanoAmoros2026, Attia2026} are not exact and can introduce density uncertainties at the level of a few percent \citep[e.g.][]{Aguichine2025, Cozza2026}. 
In addition, treating mixtures with an ideal mixing law neglects non-ideal interactions between materials. 
This simplification can further modify the resulting density profiles \citep[e.g.][]{Darafeyeu2024} and, consequently, the inferred bulk composition. 
Related to this, only a limited set of materials is included in the models. 
Specifically, the analysis considers H--He, water, rock, and iron, while other plausible constituents such as methane or ammonia are neglected. 
The latter two limitations, namely restricting the composition space and simplifying mixture behaviour, effectively reduce the number of free parameters. 
A more complete and physically complex treatment would hence maintain or even increase the degeneracy discussed in this work.
While the results presented in this work correspond to a specific choice of EoSs, the key conclusion that interior models with composition gradients are more degenerate than purely adiabatic ones is hence robust.
Finally, EoS tables can also suffer from unphysical features and interpolation issues (Fig.~\ref{fig:eos-grad-ad_H2O}).
To avoid unstable, negative adiabatic temperature gradients, we clipped
\begin{equation}
    \nabla_\mathrm{ad} \leftarrow
\begin{cases}
  0.01, & \text{if } \nabla_\mathrm{ad} < 0.01, \\
  \nabla_\mathrm{ad}, & \text{if } 0.01 \le \nabla_\mathrm{ad} \le 0.5, \\
  0.5, & \text{if } \nabla_\mathrm{ad} > 0.5,
\end{cases}
\label{eq:clipping}
\end{equation}
for all employed EoSs.
As Fig.~\ref{fig:eos-grad-ad_H2O} shows, the clipping only affects narrow regions of the EoS tables near phase transitions or table boundaries.
However, it can noticeably change the results. 
For example, not applying Eq.~\eqref{eq:clipping} can remove cold water-rich solutions as such profiles cross the solid--liquid phase transition region, pick up a negative temperature gradient, and hence are ultimately discarded by the algorithm.

%%%%%%%%%%%%%%%%%%%%%%%%%%%%%%%%%%%%%%%%%%%%%%%%%%%%%%%%%%%%%%

%%%%%%%%%%%%%%%%%%%%%%%%%%%%%%%%%%%%%%%%%%%%%%%%%%%%%%%%%%%%%%

Second, our models assume non-rotating planets, as no reliable rotation measurements are available for the considered exoplanets. 
This assumption enforces spherical symmetry and implies vanishing gravitational moments. 
In reality, rotation can alter the internal density distribution and modify the range of admissible interior structures \citep[e.g.][]{Neuenschwander2024}.

%%%%%%%%%%%%%%%%%%%%%%%%%%%%%%%%%%%%%%%%%%%%%%%%%%%%%%%%%%%%%%

Third, several structural simplifications were imposed. 
The composition was assumed to be either constant or to vary linearly as a function of radius throughout the interior. 
Furthermore, in regions with linear gradients, the temperature gradient was prescribed through Eq.~\eqref{eq:grad_T_stable}, where we adopted a constant value of $R_\rho = 0.01$.
Although this assumption is consistent with \cite{Howard2025}, it represents only one example and does not capture the full range of possible thermal profiles.  

%%%%%%%%%%%%%%%%%%%%%%%%%%%%%%%%%%%%%%%%%%%%%%%%%%%%%%%%%%%%%%

Fourth, we emphasise that the solutions shown in Figs.~\ref{fig:K2-18b_hhe-vs-rock_pure-s1-T300+T600_by-Mg2SiO4-Fe}, \ref{fig:K2-18b_hhe-vs-rock_pure+grad-s1-T300_by-Mg2SiO4-Fe}, \ref{fig:K2-18b_ternary_pure+grad-s1-T300}, \ref{fig:K2-18b_hhe-vs-rock_grad-s1-T300+T600_by-H2O-Mg2SiO4}, and \ref{fig:eos-grad-ad_H2O} reflect our sampling procedure and should not be interpreted as a probability distribution.
The minimum and maximum values listed in Table~\ref{tab:Abundances_K2-18b_300K} correspond to the boundaries of the sampled models and are therefore comparatively less sensitive to the sampling measure, although they do depend on the sampling coverage.
The Spearman rank correlation coefficients describe the properties of our particular model ensemble and therefore depend on the sampling procedure.
Therefore, the associated $p$ values should not be interpreted as evidence for statistical relations in the observed planets, but only as formal measures of the correlations within our sampled ensembles.

%%%%%%%%%%%%%%%%%%%%%%%%%%%%%%%%%%%%%%%%%%%%%%%%%%%%%%%%%%%%%%

Fifth, for the purely adiabatic models, some regions of the cold models may contain rock and iron in solid states, which could inhibit convection and therefore make our assumption of fully convective heat transport in these models less applicable.
For the composition-gradient models, the heavy components are present as mixtures, for which the phase behaviour and corresponding transport properties are uncertain and cannot be directly inferred from the melting curves of the pure components.
In all cases, however, our models imposed a minimum temperature of 2000\,K for the presence of rock and iron to avoid their existence in a solid state.
This cut does not guarantee that rock and iron are molten, as both can remain solid above 2000\,K at multi-megabar pressures, but it removes the coldest models.

%%%%%%%%%%%%%%%%%%%%%%%%%%%%%%%%%%%%%%%%%%%%%%%%%%%%%%%%%%%%%%

These limitations may bias our findings or limit the diversity of admissible interior states.
However, while these limitations affect the quantitative results, they do not weaken the qualitative conclusion that composition gradients allow for a larger variety of interior structures.

%%%%%%%%%%%%%%%%%%%%%%%%%%%%%%%%%%%%%%%%%%%%%%%%%%%%%%%%%%%%%%

Finally, this work aims to show how composition gradients reshape the space of admissible solutions. 
It is yet to be determined under what conditions composition gradients are expected in intermediate-mass planets and whether they are more common than purely adiabatic interiors. 
In order to assess this topic, further knowledge of various physical and chemical processes such as the formation history, miscibility and phase equilibria, convective mixing and semi-convection, as well as giant impacts that can affect the planetary structure is required, and we hope to address some of them in future research.

%%%%%%%%%%%%%%%%%%%%%%%%%%%%%%%%%%%%%%%%%%%%%%%%%%%%%%%%%%%%%%

\section{Conclusions}
\label{sec:Conclusions}

%%%%%%%%%%%%%%%%%%%%%%%%%%%%%%%%%%%%%%%%%%%%%%%%%%%%%%%%%%%%%%

We presented more realistic interior models of sub-Neptunes that go beyond the simple fully homogeneous and adiabatic assumption.
We computed more complex models with composition gradients that are stable against convection and consistent with recent ab initio calculations of hydrogen and water demixing \citep{Howard2025}.
By comparing these two model classes -- purely adiabatic and including composition gradients -- for K2-18\,b and TOI-270\,d, we show that composition gradients systematically expand the range of possible interior structures and bulk compositions. 
We show that if such models are considered, improving the uncertainties of the mass and radius measurements or atmospheric boundary conditions does not necessarily decrease the degeneracy in the inferred internal structure and composition.
As a result, the actual level of degeneracy in exoplanet interior modelling is even larger than typically assumed. 
Our main conclusions can be summarised as follows: 

%%%%%%%%%%%%%%%%%%%%%%%%%%%%%%%%%%%%%%%%%%%%%%%%%%%%%%%%%%%%%%

\begin{itemize}
    \setlength{\itemsep}{0.5em}
    
    \item Degeneracy is an inherent property:
    the difficulty of uniquely inferring planetary interiors is primarily due to the complex physics and model assumptions in planetary models.
    Improving observational constraints, such as planetary mass and radius, can therefore only marginally reduce this inherent degeneracy.

    \item Adiabatic models underestimate the possible solution space:
    purely adiabatic models can restrict the physically allowed solution space.
    Allowing for composition gradients reveals a much wider range of possible interiors, including substantially higher H--He mass fractions, by a factor of about three at a 1\,bar temperature of 300\,K and by a factor of about 10 at 600\,K for K2-18\,b and, generally, by more than an order of magnitude for TOI-270\,d.

    \item Composition gradients introduce additional uncertainty:
    correlations are often weaker in models with composition gradients than in purely adiabatic models.
    For example, the Spearman rank correlation coefficient between the H--He and iron abundances decreases from $\sim0.7$ to $\sim0.3$ for both K2-18\,b and TOI-270\,d.
    Rock-to-iron ratios also become less correlated with the planetary H--He content.
    External constraints, such as host-star abundance ratios, should therefore be used with caution, as their validity depends on the assumptions of the interior model.

    \item External constraints remain useful:
    composition gradients do not eliminate correlations.
    Some correlations, such as that between the H--He abundance and the ice-to-rock ratio, can persist or even strengthen and remain non-negligible.
    Independent constraints from planet formation, migration, or protoplanetary disc chemistry can therefore still help narrow down the possible interior structures and bulk compositions of planets.
    
\end{itemize}

%%%%%%%%%%%%%%%%%%%%%%%%%%%%%%%%%%%%%%%%%%%%%%%%%%%%%%%%%%%%%%

We hence suggest that interior models used for exoplanet characterisation should consider mixtures of materials, composition gradients, and non-convective regions by default in order to capture the full range of physically plausible solutions. 
For sub-Neptunes, the question is not whether the interiors are complex, but how much complexity current models are still missing. 
Composition gradients are therefore not an optional detail but part of the answer. 
Such structures are an expected outcome of planet formation and evolution simulations \citep{Helled2017, Valletta2022, Eberlein2025, Eberlein2026} and could also emerge due to miscibility \citep{Young2024}.
Neglecting them can bias our conclusions regarding the composition and potential habitability of exoplanets.

%%%%%%%%%%%%%%%%%%%%%%%%%%%%%%%%%%%%%%%%%%%%%%%%%%%%%%%%%%%%%%

Looking ahead, the coming decade promises transformative progress in our understanding of sub-Neptunes. 
The growing synergy between space-based missions and increasingly capable ground-based observatories will provide unprecedented constraints on the physical and chemical properties of these worlds across a wide range of environments. 
At the same time, advancements in numerical simulations and theoretical modelling will enable more realistic interpretations of the expanding observational dataset, helping to connect planetary formation, evolution, and atmospheric processes within a unified framework. 
By combining these complementary approaches, the field is poised to move beyond population-level trends towards a deeper, more predictive understanding of the origin and diversity of sub-Neptunes.

%%%%%%%%%%%%%%%%%%%%%%%%%%%%%%%%%%%%%%%%%%%%%%%%%%%%%%%%%%%%%%

\begin{acknowledgements}

    We thank Simon M\"uller for providing the EoS code and valuable discussions. 
    We also thank the referee for helpful suggestions. 
    This work was supported by the Swiss National Science Foundation (SNSF) via grant number 215634: \url{https://data.snf.ch/grants/grant/215634}.
      
\end{acknowledgements}

%%%%%%%%%%%%%%%%%%%%%%%%%%%%%%%%%%%%%%%%%%%%%%%%%%%%%%%%%%%%%%

\bibliographystyle{aa}
\bibliography{literature}

%%%%%%%%%%%%%%%%%%%%%%%%%%%%%%%%%%%%%%%%%%%%%%%%%%%%%%%%%%%%%%%

\begin{appendix}
\nolinenumbers

%%%%%%%%%%%%%%%%%%%%%%%%%%%%%%%%%%%%%%%%%%%%%%%%%%%%%%%%%%%%%%%

\section{Additional figures for \texorpdfstring{TOI-270\,d}{TOI-270 d}}
\label{sec:further_figures}

%%%%%%%%%%%%%%%%%%%%%%%%%%%%%%%%%%%%%%%%%%%%%%%%%%%%%%%%%%%%%%%

\begin{figure}
    \centering
    \includegraphics[width=\linewidth]{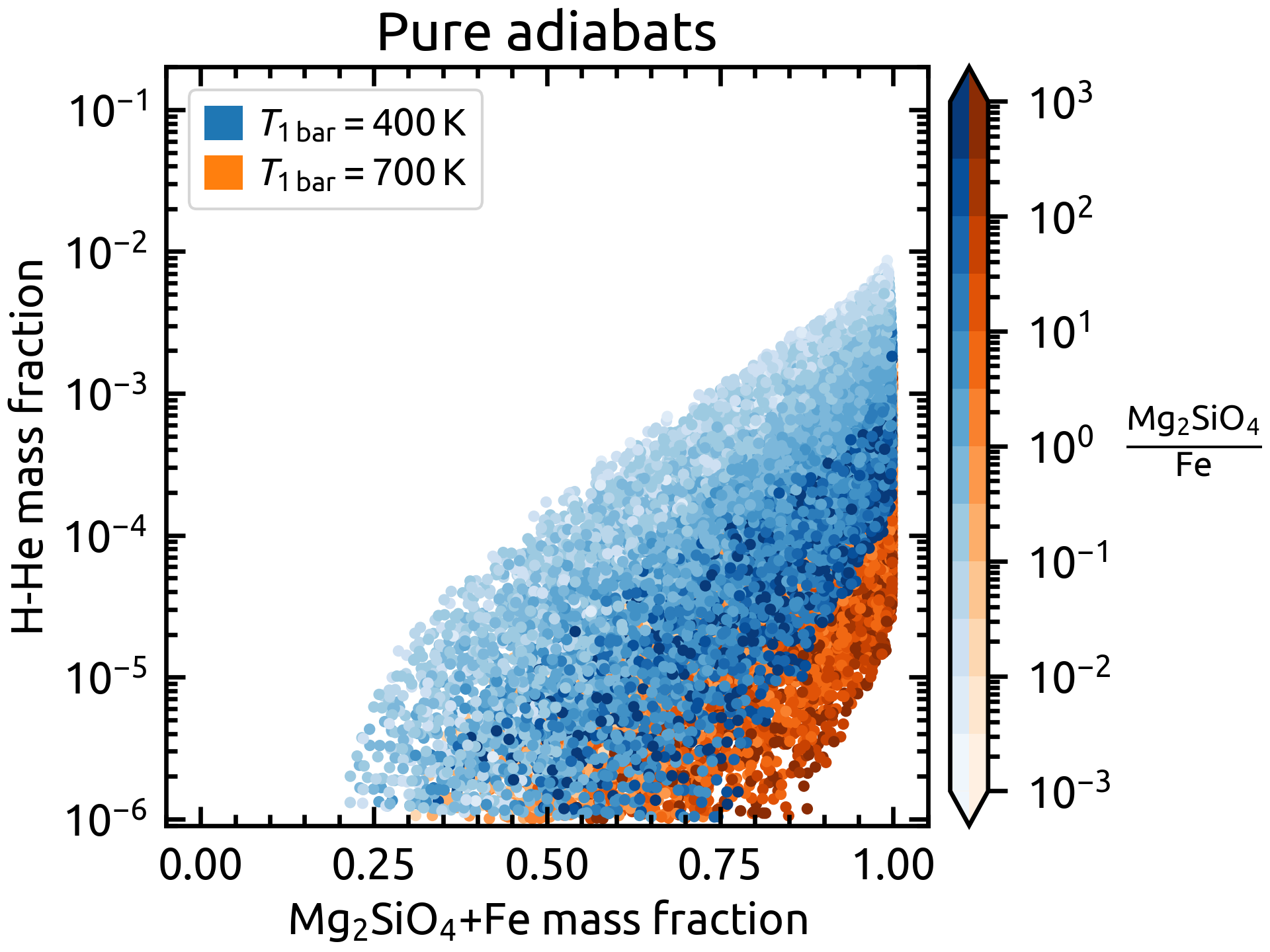}
    \caption{
    Same as Fig.~\ref{fig:K2-18b_hhe-vs-rock_pure-s1-T300+T600_by-Mg2SiO4-Fe} but for TOI-270\,d. 
    }
    \label{fig:TOI-270d_hhe-vs-rock_pure-s1-T400+T700_by-Mg2SiO4-Fe}
\end{figure}

\begin{figure}
    \includegraphics[width=\linewidth]{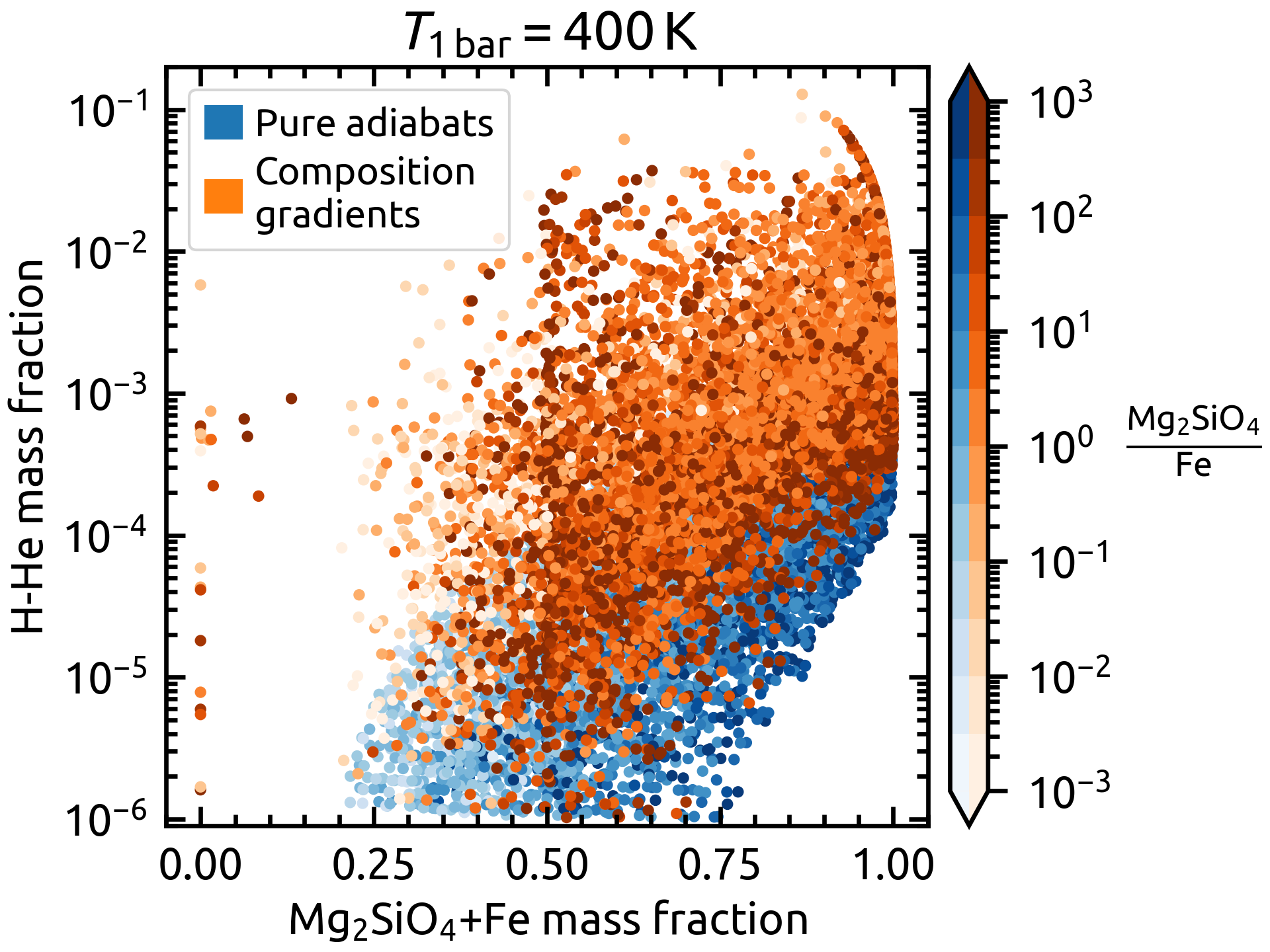}
    \caption{
    Same as Fig.~\ref{fig:K2-18b_hhe-vs-rock_pure+grad-s1-T300_by-Mg2SiO4-Fe} but for TOI-270\,d.
    }
    \label{fig:TOI-270d_hhe-vs-rock_pure+grad-s1-T400_by-Mg2SiO4-Fe}
\end{figure}

\begin{figure}
    \includegraphics[width=\linewidth]{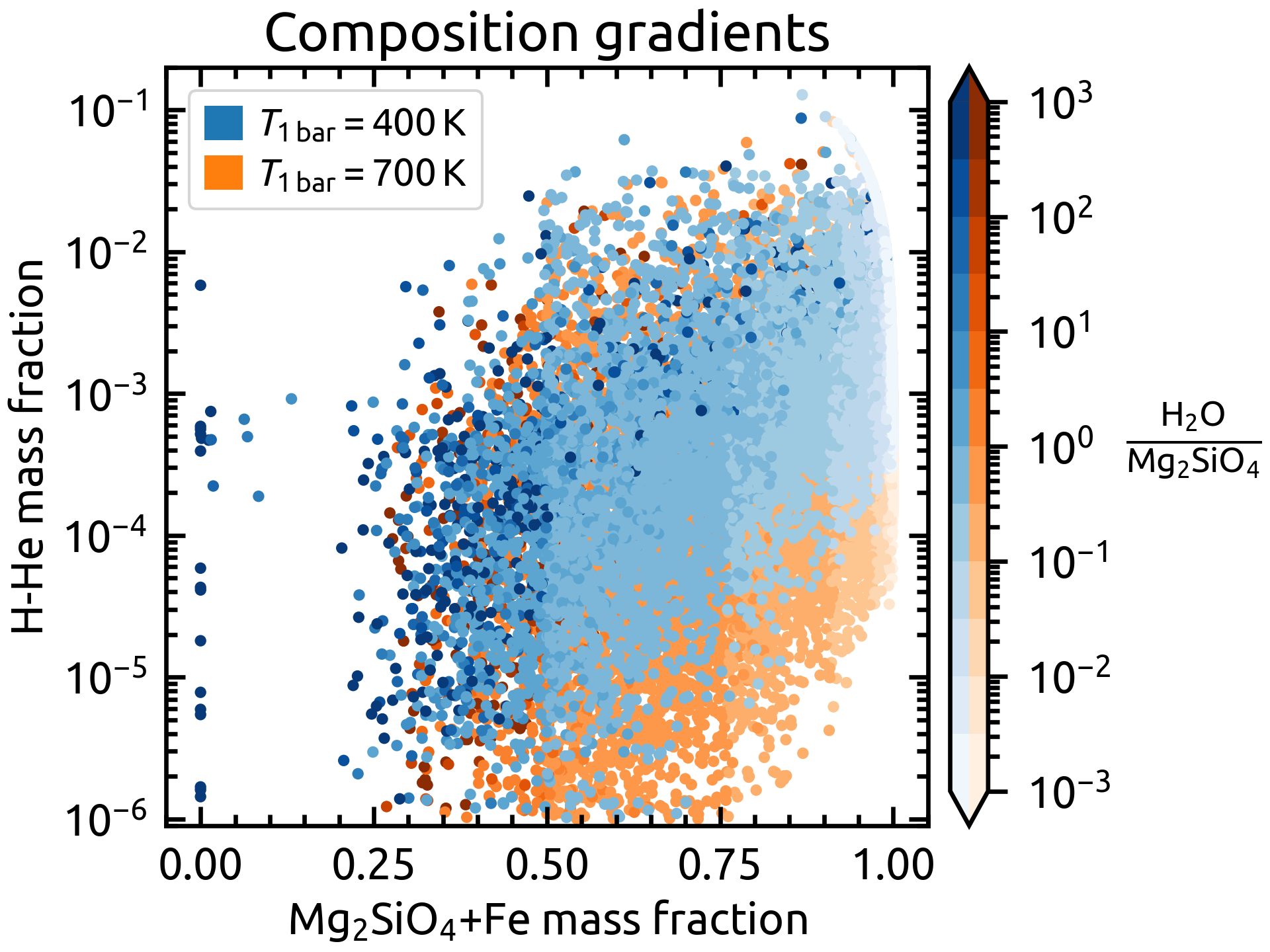}
    \caption{
    Same as Fig.~\ref{fig:K2-18b_hhe-vs-rock_grad-s1-T300+T600_by-H2O-Mg2SiO4} but for TOI-270\,d.
    }
    \label{fig:TOI-270d_hhe-vs-rock_grad-s1-T400+T700_by-H2O-Mg2SiO4}
\end{figure}

\begin{figure*}
    \centering
    \includegraphics[width=0.49\linewidth]{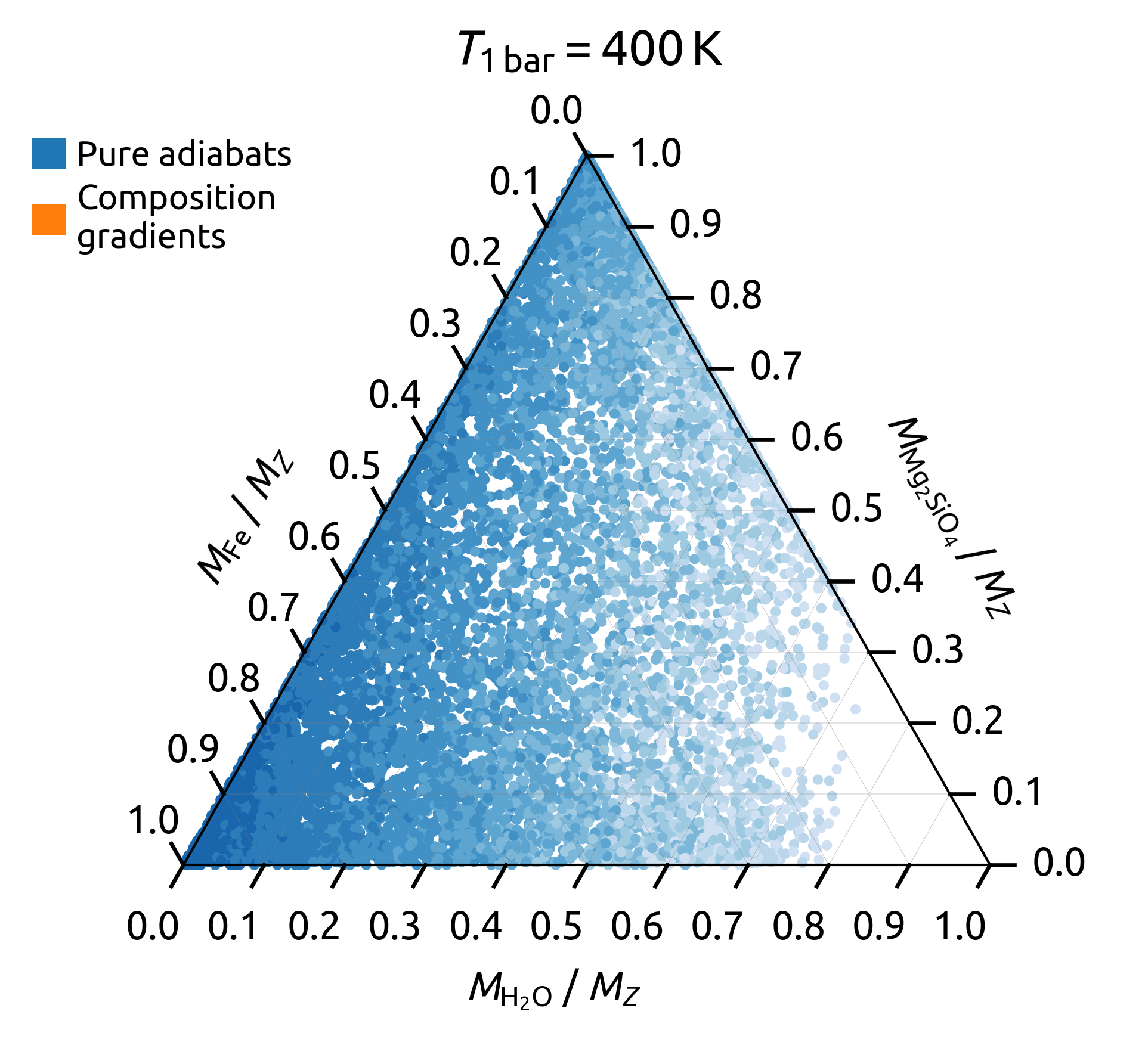}
    \includegraphics[width=0.49\linewidth]{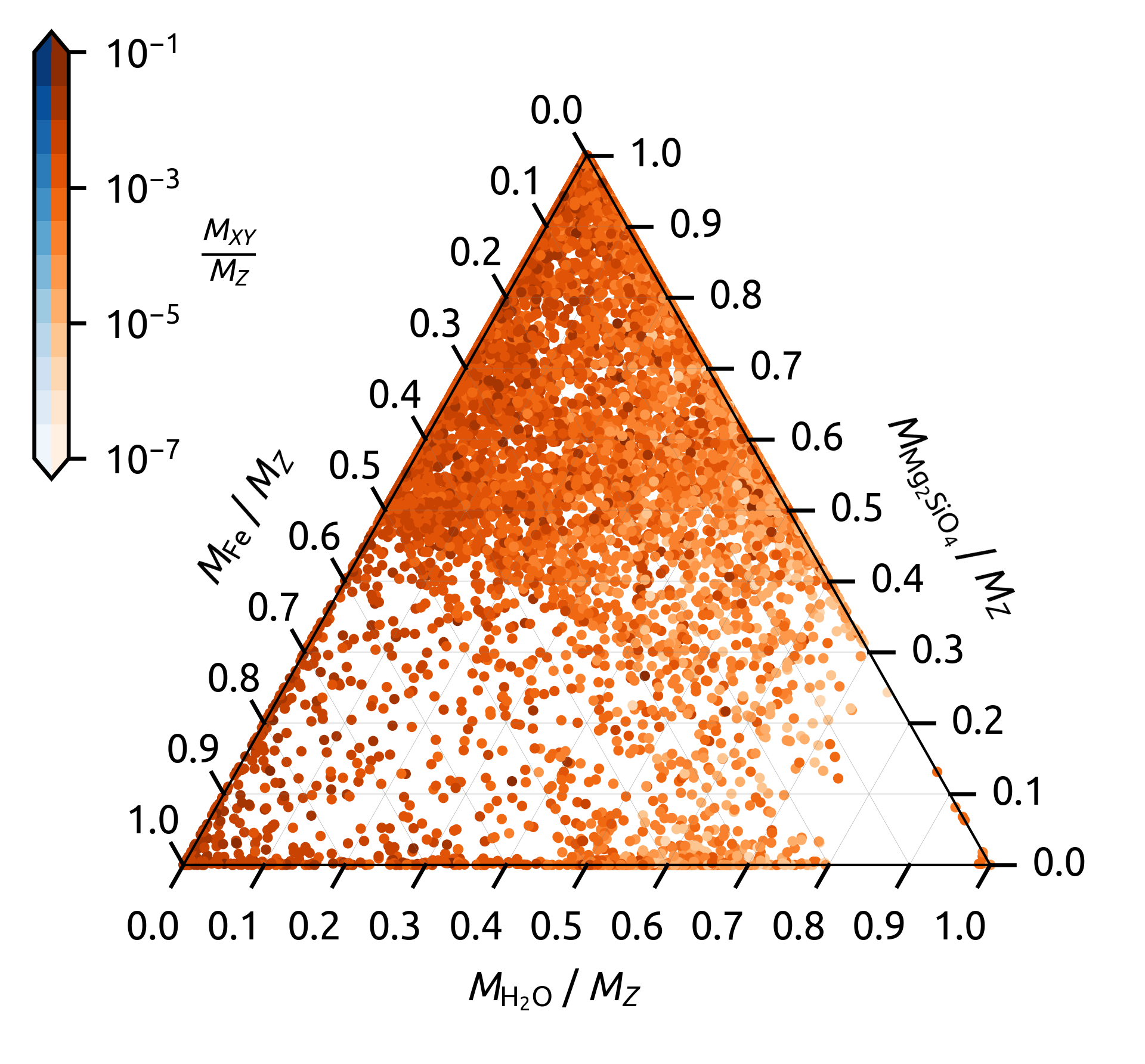}
    \caption{
    Same as Fig.~\ref{fig:K2-18b_ternary_pure+grad-s1-T300} but for TOI-270\,d.
    }
    \label{fig:TOI-270d_ternary_pure+grad-s1-T400}
\end{figure*}

\begin{figure}
    \centering
    \includegraphics[width=\linewidth]{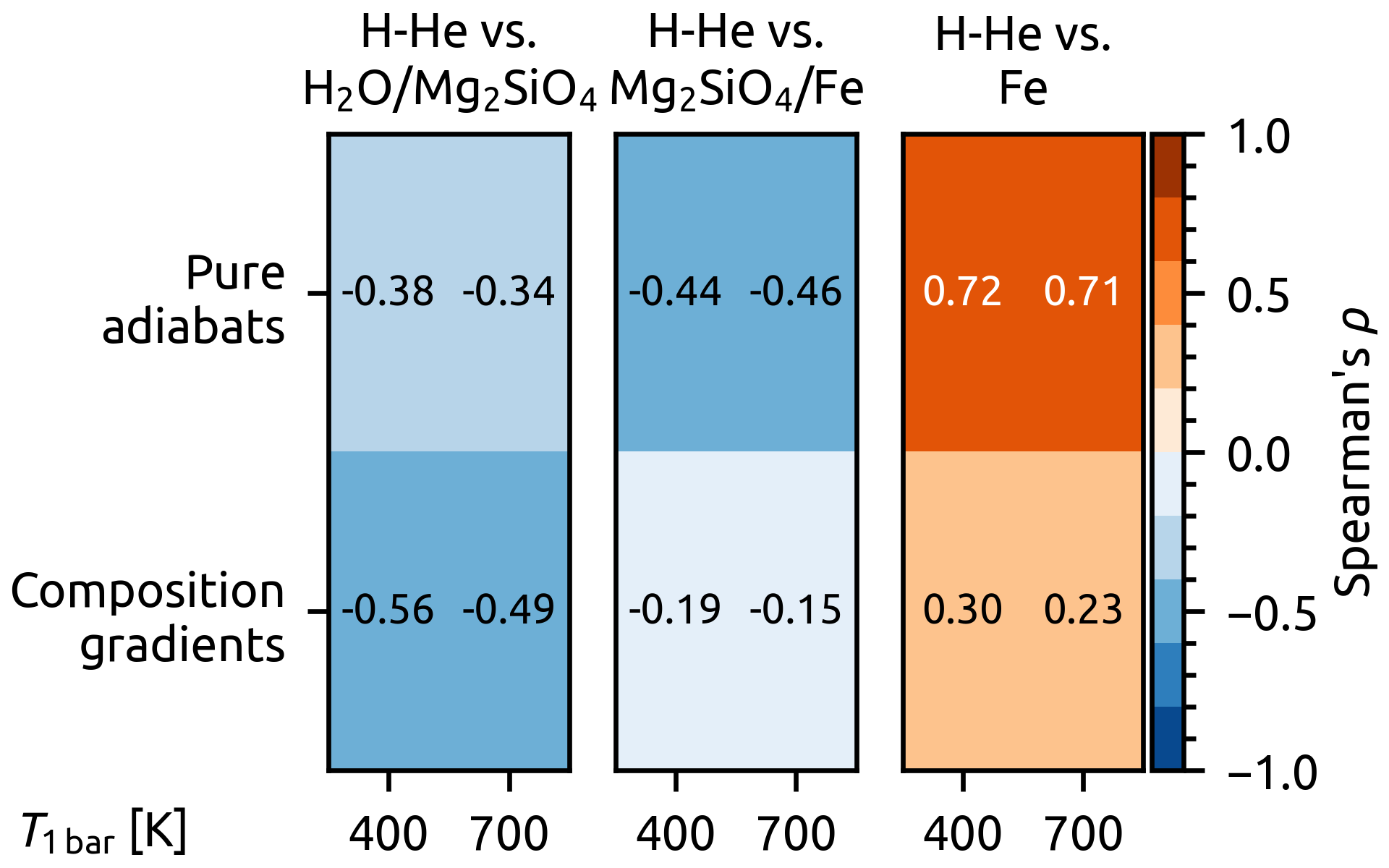}
    \caption{
    Same as Fig.~\ref{fig:K2-18b_corr} but for TOI-270\,d. 
    The inferred $p$ values, which are smaller than $10^{-52}$, are valid only within our modelling framework (Sect.~\ref{sec:Limitations}).
    }
    \label{fig:TOI-270d_corr}
\end{figure}

Table~\ref{tab:Abundances_TOI-270d_400K} and Figs.~\ref{fig:TOI-270d_hhe-vs-rock_pure-s1-T400+T700_by-Mg2SiO4-Fe}, \ref{fig:TOI-270d_hhe-vs-rock_pure+grad-s1-T400_by-Mg2SiO4-Fe}, \ref{fig:TOI-270d_hhe-vs-rock_grad-s1-T400+T700_by-H2O-Mg2SiO4}, \ref{fig:TOI-270d_ternary_pure+grad-s1-T400}, and \ref{fig:TOI-270d_corr} are similar to the results presented in Sect.~\ref{sec:Results} but for TOI-270\,d. 
We include these to demonstrate that our results are not valid just for a specific planet, but rather apply to all the observed sub-Neptunes.

%%%%%%%%%%%%%%%%%%%%%%%%%%%%%%%%%%%%%%%%%%%%%%%%%%%%%%%%%%%%%%%

Table~\ref{tab:Abundances_TOI-270d_400K} displays the low-temperature ($T_{1\,\mathrm{bar}}=400\,$K) case. 
We find that purely adiabatic models only allow for negligible amounts of H--He below or equal to 1\%, while for more complex models (with composition gradients) the inferred H--He mass fractions can be up to 13\%. 
We refer the reader to Sect.~\ref{sec:Results} for details. 

%%%%%%%%%%%%%%%%%%%%%%%%%%%%%%%%%%%%%%%%%%%%%%%%%%%%%%%%%%%%%%%

Figures~\ref{fig:TOI-270d_hhe-vs-rock_pure+grad-s1-T400_by-Mg2SiO4-Fe} and \ref{fig:TOI-270d_hhe-vs-rock_grad-s1-T400+T700_by-H2O-Mg2SiO4} show that our sampling procedure is able to generate nearly pure water composition-gradient models (with a vanishing combined rock and iron mass fraction) at a very low frequency.
Such models are missing entirely for analogous purely adiabatic models.

%%%%%%%%%%%%%%%%%%%%%%%%%%%%%%%%%%%%%%%%%%%%%%%%%%%%%%%%%%%%%%%

Figure~\ref{fig:TOI-270d_corr} shows that for TOI-270\,d, the same conclusion holds as for K2-18\,b: 
the H--He versus Mg$_2$SiO$_4$/Fe and H--He versus Fe correlations are weaker in models with composition gradients than in purely adiabatic models.
In contrast, the H--He versus H$_2$O/Mg$_2$SiO$_4$ correlations remain significant and even increase moderately.

%%%%%%%%%%%%%%%%%%%%%%%%%%%%%%%%%%%%%%%%%%%%%%%%%%%%%%%%%%%%%%%

\begin{table}
\caption{
Same as Table~\ref{tab:Abundances_K2-18b_300K} but for TOI-270\,d.
}
\centering
\setlength{\tabcolsep}{5.5pt}
    \begin{tabular}{lllcccc}
    \hline
    \hline
    TOI-270\,d & $\sigma_{R},\sigma_{M}$ & $T_{1\,\mathrm{bar}}$ & \multicolumn{2}{c}{\tiny{H--He [\%]}} & \multicolumn{2}{c}{\tiny{Mg$_2$SiO$_4$+Fe [\%]}} \\
    \cline{4-5} \cline{6-7}
    & & & \tiny{min} & \tiny{max} & \tiny{min} & \tiny{max} \\
    \hline \hline
    \multirow{2}{*}{\makecell[l]{Pure \\ adiabats}} & =0 & 400\,K & 0 & 1 & 33 & 100 \\
    & >0 & 400\,K & 0 & 1 & 22 & 100 \\
    \hline
    \multirow{2}{*}{\makecell[l]{Composition \\ gradients}} & =0 & 400\,K & 0 & 4 & 6 & 100 \\
    & >0 & 400\,K & 0 & 13 & 0 & 100 \\
    \hline \hline
    Ben+24 & >0 & & 0 & 1 & 45 & 80 \\
    \hline
    How+25 & >0 & & 4 & 4 & 90 & 90 \\
    \hline
    Rig+26 & >0 & & 0 & 5 & -- & 98 \\
    \hline \hline
    \end{tabular}
\label{tab:Abundances_TOI-270d_400K}
\tablefoot{
For comparison, the findings by \cite{Benneke2024, Howard2025, Rigby2026} are included.
}
\end{table}

%%%%%%%%%%%%%%%%%%%%%%%%%%%%%%%%%%%%%%%%%%%%%%%%%%%%%%%%%%%%%%%

\section{Hydrogen and rock miscibility}
\label{sec:hydro_rock_miscibility}

\begin{figure}[!t]
    \centering
    \includegraphics[width=\linewidth]{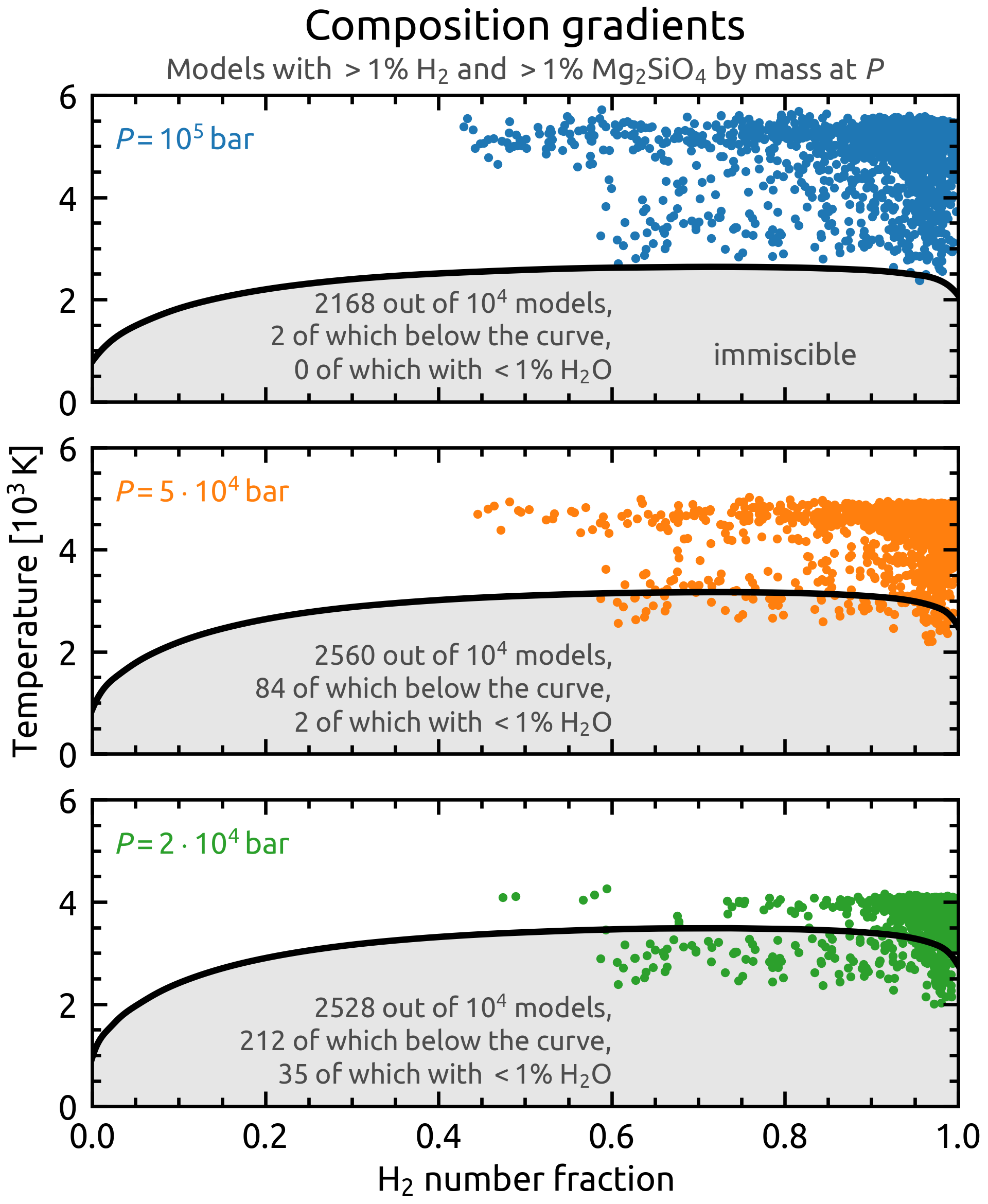}
    \caption{
    Miscibility curves (lines) by \cite{Gilmore2026} for a hydrogen fluid and a liquid of MgSiO$_3$ composition.
    The blue, orange, and green colours indicate a pressure of 10$^5$, $5\cdot10^4$, and $2\cdot10^4$\,bar, respectively.
    Each dot shows a composition-gradient model ($\sigma_R,\sigma_M>0$ case) of K2-18\,b with a 1\,bar temperature of 300\,K and non-negligible amounts of hydrogen and rock.
    Models above their respective miscibility curve can be considered stable against demixing.
    }
    \label{fig:K2-18b_miscibility-scatter_grad-s1-T300}
\end{figure}

%%%%%%%%%%%%%%%%%%%%%%%%%%%%%%%%%%%%%%%%%%%%%%%%%%%%%%%%%%%%%%%

Our models already enforce hydrogen and water miscibility according to \cite{Howard2025}.
In this section, we also consider hydrogen--rock miscibility based on the recent results of \cite{Gilmore2026}.
Figure~\ref{fig:K2-18b_miscibility-scatter_grad-s1-T300} compares their miscibility curves with $10^4$ composition-gradient models of K2-18\,b ($\sigma_R,\sigma_M>0$ case), assuming a 1\,bar temperature of 300\,K.

%%%%%%%%%%%%%%%%%%%%%%%%%%%%%%%%%%%%%%%%%%%%%%%%%%%%%%%%%%%%%%%

We find that the clear majority of our models that possess more than 1\% H--He and rock by mass for a given pressure lie above the corresponding miscibility curves.
This suggests that the inferred structures are stable against hydrogen--rock demixing. 
The others may undergo demixing, rendering them potentially unstable. 
However, our results are only qualitative:
First, we used Mg$_2$SiO$_4$ whereas \cite{Gilmore2026} consider MgSiO$_3$.
Second, our models also contain water and iron, which are not considered in \cite{Gilmore2026}.
Such additional materials can substantially alter the underlying chemistry and miscibility behaviour.
In fact, only a negligible subset of 35 models (0.35\% of all models) possess both less than 1\% water by mass and lie below the demixing line at $P=2\cdot10^4$\,bar.
We encourage further calculations and experiments of demixing in planetary conditions that involve various mixtures.

%%%%%%%%%%%%%%%%%%%%%%%%%%%%%%%%%%%%%%%%%%%%%%%%%%%%%%%%%%%%%%

\section{Sensitivity test for \texorpdfstring{$R_\rho$}{R\_rho}}
\label{sec:sens_test_Rp}

%%%%%%%%%%%%%%%%%%%%%%%%%%%%%%%%%%%%%%%%%%%%%%%%%%%%%%%%%%%%%%%

\begin{figure}
    \centering
    \includegraphics[width=\linewidth]{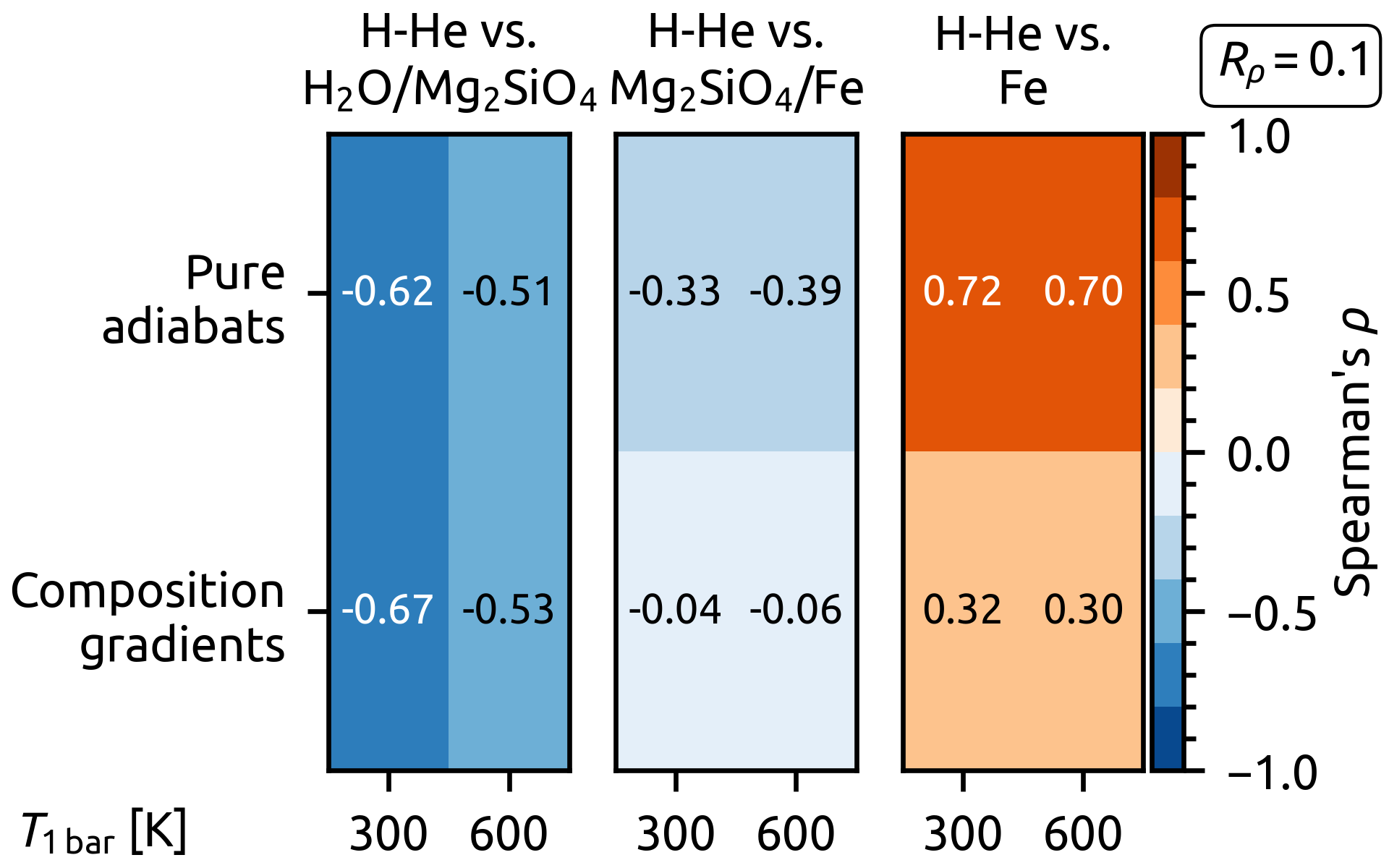}
    \caption{
    Same as Fig.~\ref{fig:K2-18b_corr} but for $R_\rho=0.1$.
    The inferred $p$ values, which are smaller than $10^{-3}$, are valid only within our modelling framework (Sect.~\ref{sec:Limitations}).
    }
    \label{fig:K2-18b_Rrho0.1_corr}
\end{figure}

As discussed in Sect.~\ref{sec:Methods}, we used $R_\rho=0.01$ to calculate the temperature gradient in regions stable against convection in our models (see Eq.~\eqref{eq:grad_T_stable}).
However, the expected value of $R_\rho$ in planetary interiors remains uncertain.
In this section, we consider models when assuming $R_\rho=0.1$.
The results are shown in Fig.~\ref{fig:K2-18b_Rrho0.1_corr}.
The correlations generally remain close to their previous values. 
Hence, composition-gradient models still show correlations that are similar to or weaker than those of the purely adiabatic models.
We can therefore conclude that our findings are robust and do not depend on the choice of $R_\rho$.
Similarly, the ranges given in Table~\ref{tab:Abundances_K2-18b_300K} are mostly unaffected:
for the $\sigma_R,\sigma_M>0$ case and purely adiabatic models, the minima and maxima are now 0\% and 2\% for the total H--He abundance and 0\% and 100\% for the total rock and iron abundance.
The analogous percentages for the composition-gradient models are 0\% and 10\%, and 0\% and 100\%, respectively.
These percentages differ only by 1\% or less from the values given in Table~\ref{tab:Abundances_K2-18b_300K}.

%%%%%%%%%%%%%%%%%%%%%%%%%%%%%%%%%%%%%%%%%%%%%%%%%%%%%%%%%%%%%%%

%%%%%%%%%%%%%%%%%%%%%%%%%%%%%%%%%%%%%%%%%%%%%%%%%%%%%%%%%%%%%%%

\section{Sampling methodology and analysis}
\label{sec:sampling_methods}

%%%%%%%%%%%%%%%%%%%%%%%%%%%%%%%%%%%%%%%%%%%%%%%%%%%%%%%%%%%%%%%

Because of the structural simplifications described in Sect.~\ref{sec:Methods}, a random density function $\rho(r)$ can be represented by only a few unconstrained parameters:
\begin{itemize}
    \item one parameter $u_R$ for the outer radius
    \item three parameters $u_{r_1}$, $u_{r_2}$, and $u_{r_3}$ for the transition radii
\end{itemize}
These parameters are sufficient for purely adiabatic models.
However, structures with composition gradients require additional parameters:
\begin{itemize}
    \item one parameter $u_\mathrm{conv}$ to determine whether the outermost layer is convective or stable
    \item fifteen parameters $u_{r}^{XY}$, $u_{r}^{Z_1}$, and $u_{r}^{Z_2}$, with $r\in\{0,r_1,r_2,r_3,R\}$, to specify the composition of each layer
\end{itemize}
We converted these unconstrained parameters into physical quantities using the sigmoid function $\operatorname{sigmoid}(x)=1/\left(1+\exp(-x)\right)$.
The outer radius $R$ is given by
\begin{equation}
R = (R_\mathrm{observation}-\sigma_R) + 2\sigma_R\cdot\operatorname{sigmoid}(u_R),
\end{equation}
which constrains $R$ to the interval $(R_\mathrm{observation}-\sigma_R, R_\mathrm{observation}+\sigma_R)$.
We set $\sigma_R=2\sigma_\mathrm{observation}$ when $\sigma_R,\sigma_M>0$.
For $\sigma_R,\sigma_M=0$, we used $\sigma_R=0.01\sigma_\mathrm{observation}$.

%%%%%%%%%%%%%%%%%%%%%%%%%%%%%%%%%%%%%%%%%%%%%%%%%%%%%%%%%%%%%%%

All model types are further defined by three transition radii:
\begin{align}
r_1/R&=\operatorname{sigmoid}(u_{r_1}), \\
r_2/R&=\operatorname{sigmoid}(u_{r_1})\cdot\operatorname{sigmoid}(u_{r_2}), \\
r_3/R&=\operatorname{sigmoid}(u_{r_1})\cdot\operatorname{sigmoid}(u_{r_2})\cdot\operatorname{sigmoid}(u_{r_3}).
\end{align}
This parametrisation guarantees $1>r_1/R>r_2/R>r_3/R>0$.
The three transition radii therefore define four distinct layers.

%%%%%%%%%%%%%%%%%%%%%%%%%%%%%%%%%%%%%%%%%%%%%%%%%%%%%%%%%%%%%%%

The procedure to generate models is now as follows.
We started with an initial pressure array $P^0=[P^0_0,\dots,P^0_{N-1}]$, where
\begin{equation}
P^0_i = 1\,\text{bar}\cdot\left(\frac{10^{13}\,\text{bar}}{1\,\text{bar}}\right)^{\left(\frac{i}{N-1}\right)^{\frac{1}{4}}}.
\end{equation}
We used $N=512$ grid points.
$T_{1\,\mathrm{bar}}$ provides the outer boundary condition.
If the composition is known, we can then integrate inwards using the EoS, assuming an adiabatic temperature gradient in convective regions.
For stable regions, we used Eq.~\eqref{eq:grad_T_stable}.
This integration yields temperature and density profiles.
The density profile in turn determines the gravitational potential, which can be used to obtain an updated pressure array $P^1$ through Eq.~\eqref{eq:HE}.
This procedure is repeated iteratively.
Convergence after $n$ iterations is reached when
\begin{equation}
\max_{i\in\{0,\dots,N-1\}}\left|\frac{P^{n+1}_i}{P^n_i}-1\right|< \varepsilon=0.02.
\end{equation}

%%%%%%%%%%%%%%%%%%%%%%%%%%%%%%%%%%%%%%%%%%%%%%%%%%%%%%%%%%%%%%%

The procedure above fully specifies the purely adiabatic case, as the three transition radii $r_1$, $r_2$, and $r_3$ uniquely determine the composition profile.
No additional parameters are required.
However, composition-gradient models require additional information:

%%%%%%%%%%%%%%%%%%%%%%%%%%%%%%%%%%%%%%%%%%%%%%%%%%%%%%%%%%%%%%%

First, we note that the same four-layer structure is retained for composition-gradient models.
This ensures a consistent comparison between the two model classes.
To determine the stability of the outermost layer, $r\in[R,r_1]$, we used the parameter $u_\mathrm{conv}$:
\begin{equation}
\text{model starts}
\begin{cases}
\text{convective}, & \text{if }\operatorname{sigmoid}(u_\mathrm{conv})>0.5,\\
\text{stable}, & \text{otherwise}.
\end{cases}
\end{equation}
The state of each subsequent layer is then obtained by alternating between convective and stable.
The four layers therefore always contain two convective and two stable regions for composition-gradient models.

%%%%%%%%%%%%%%%%%%%%%%%%%%%%%%%%%%%%%%%%%%%%%%%%%%%%%%%%%%%%%%%

Second, we note that we assumed that the composition was constant within convective layers and varied linearly within stable layers.
Therefore, the composition is completely determined by fixing the values at each boundary $r\in\{0,r_1,r_2,r_3,R\}.$
The four mass fractions defined at each boundary correspond to H--He, water, rock, and iron.
We denote them by $XY_r$, $Z_{1,r}$, $Z_{2,r}$, and $Z_{3,r}$, respectively.
We obtained these fractions using a softmax-like transformation:
\begin{align}
XY_r &= \frac{\exp\left(u_r^{XY}/\tau\right)}{\mathcal{N}_r}, &&Z_{1,r} = \frac{\exp\left(u_r^{Z_1}/\tau\right)}{\mathcal{N}_r}, \label{eq:E1} \\
Z_{2,r} &= \frac{\exp\left(u_r^{Z_2}/\tau\right)}{\mathcal{N}_r},
&&Z_{3,r} = \frac{1}{\mathcal{N}_r}, \label{eq:E2}
\end{align}
where
\begin{equation}
\mathcal{N}_r = 1+\exp\left(u_r^{XY}/\tau\right)+\exp\left(u_r^{Z_1}/\tau\right)+\exp\left(u_r^{Z_2}/\tau\right), \label{eq:E3}
\end{equation}
for every $r\in\{0,r_1,r_2,r_3,R\}.$
The parameter $\tau$ controls the sharpness of the transformation.
Its role is discussed below.
Note that this parametrisation has two useful properties.
First, $XY_r+Z_{1,r}+Z_{2,r}+Z_{3,r}=1$ by construction.
Second, all mass fractions are strictly positive: $XY_r,Z_{1,r},Z_{2,r},Z_{3,r}>0$.

%%%%%%%%%%%%%%%%%%%%%%%%%%%%%%%%%%%%%%%%%%%%%%%%%%%%%%%%%%%%%%%

Regarding the prior distributions, we adopted:
\begin{align}
u_R &\sim \operatorname{Logistic}(0,1), 
\\
\begin{pmatrix}u_{r_1}\\u_{r_2}\\u_{r_3}\end{pmatrix} &\sim\operatorname{Logistic}
\left[\begin{pmatrix}1\\1\\1\end{pmatrix},1\right], \label{eq:E5} 
\\
u_{\mathrm{conv}} &\sim \operatorname{Logistic}(0,1),
\\
\begin{pmatrix}
u_R^{XY} & u_R^{Z_1} & u_R^{Z_2}\\
u_{r_1}^{XY} & u_{r_1}^{Z_1} & u_{r_1}^{Z_2}\\
u_{r_2}^{XY} & u_{r_2}^{Z_1} & u_{r_2}^{Z_2}\\
u_{r_3}^{XY} & u_{r_3}^{Z_1} & u_{r_3}^{Z_2}\\
u_0^{XY} & u_0^{Z_1} & u_0^{Z_2}
\end{pmatrix}
&\sim \operatorname{Logistic} 
\left[\begin{pmatrix}
3 & 1 & -1\\
1 & 1 & -1\\
0 & 1 & 1\\
-1 & 0 & 1\\
-3 & -1 & -1
\end{pmatrix},1\right]. \label{eq:E4}
\end{align}
$\operatorname{Logistic}(\mu,s)$ denotes the logistic distribution with mean $\mu$ and scale parameter $s$.
In particular, if $Y\sim\operatorname{Logistic}(0,1)$, then
\begin{equation}
    \operatorname{sigmoid}(Y)\sim\operatorname{Uniform}(0,1),
\end{equation}
which motivates the use of logistic priors.
It produces uniform priors for $\operatorname{sigmoid}(u_R)$ and $\operatorname{sigmoid}(u_{\mathrm{conv}})$.
For the transition radius parameters we used a mean of $1$ instead of $0$.
This shifts their distribution towards larger values, ensuring sufficient sampling of configurations with $r_3$ close to $R$.
For the composition parameters we chose means such that lighter materials are favoured in the outer layers and heavier materials are favoured at greater depths.
Finally, note that we set $\tau=0.2$ to ensure that the softmax transformation covers a broad range of possible composition gradients, including pure end-members where one component dominates strongly. 
The former two choices reflect the behaviour of the purely adiabatic models and therefore allow for a fairer comparison between the two.

\begin{figure}
    \centering
    \includegraphics[width=\linewidth]{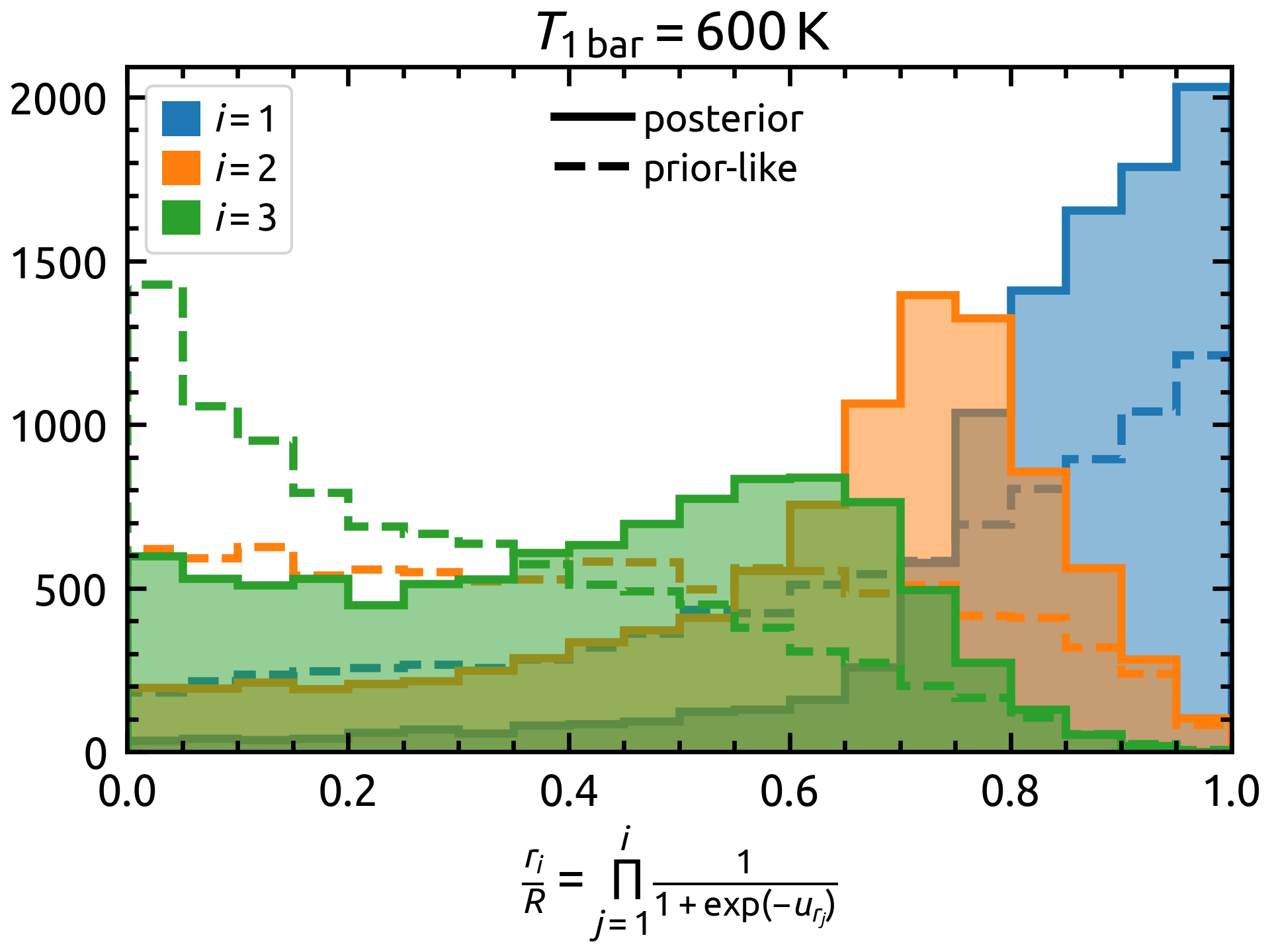}
    \caption{
    Histogram of the normalised transition radii $r_1/R$, $r_2/R$, and $r_3/R$.
    Shown are $10^4$ K2-18\,b composition-gradient models for the $\sigma_R,\sigma_M>0$ case with $T_{1\,\mathrm{bar}}=600\,$K.
    The solid lines show the posteriors consistent with the observed mass, while the dashed lines show a prior-like distribution as shown in Eq.~\eqref{eq:E5}.
    }
    \label{fig:K2-18b_posterior-us_grad-s1-T600}
\end{figure}

%%%%%%%%%%%%%%%%%%%%%%%%%%%%%%%%%%%%%%%%%%%%%%%%%%%%%%%%%%%%%%%

For the $\sigma_R,\sigma_M>0$ case, we sampled the parameters directly from their prior distributions, retaining only models whose mass $M$ satisfies $\left|M-M_{\mathrm{observation}}\right|<2\sigma_{\mathrm{observation}}$.
For the $\sigma_R,\sigma_M=0$ case, the required mass tolerance is much narrower, $\left|M-M_{\mathrm{observation}}\right|<0.01\sigma_{\mathrm{observation}}$.
We therefore used an MCMC search \citep{ForemanMackey2013} to more quickly find solutions.
Note that all presented correlation results are based on the $\sigma_R,\sigma_M>0$ case and hence are not affected by the MCMC sampling procedure used for the $\sigma_R,\sigma_M=0$ case.

%%%%%%%%%%%%%%%%%%%%%%%%%%%%%%%%%%%%%%%%%%%%%%%%%%%%%%%%%%%%%%%

\begin{figure}
    \centering
    \includegraphics[width=\linewidth]{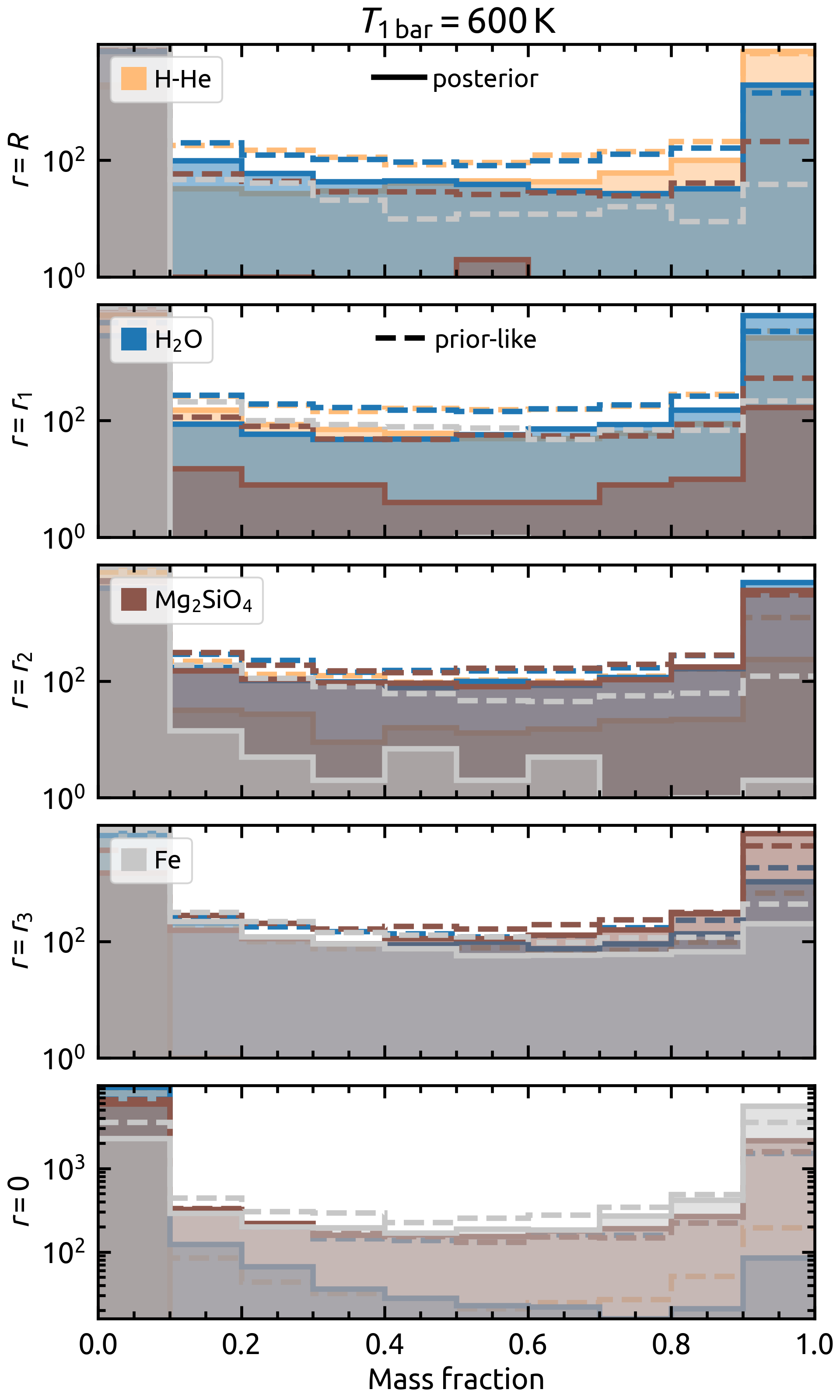}
    \caption{
    Histograms of the mass fractions $XY_r$, $Z_{1,r}$, $Z_{2,r}$, and $Z_{3,r}$ for every $r\in\{0,r_1,r_2,r_3,R\}$.
    They represent the mass fractions of H--He, water, rock, and iron, respectively, and were calculated from the unconstrained parameters shown in Eqs.~\eqref{eq:E1}, \eqref{eq:E2}, and \eqref{eq:E3}.
    We show the same $10^4$ models as in Fig.~\ref{fig:K2-18b_posterior-us_grad-s1-T600}.
    The solid lines show the posteriors consistent with the observed mass, while the dashed lines show a prior-like distribution as shown in Eq.~\eqref{eq:E4}.
    }
    \label{fig:K2-18b_posterior-fractions_grad-s1-T600}
\end{figure}

Figures~\ref{fig:K2-18b_posterior-us_grad-s1-T600} and \ref{fig:K2-18b_posterior-fractions_grad-s1-T600} show the prior and posterior distributions of the transition radius and composition parameters.
We considered the $\sigma_R,\sigma_M>0$ case with composition gradients for K2-18\,b and used $T_{1\,\mathrm{bar}}=600\,$K.
We do not show figures for $\operatorname{sigmoid}(u_R)$ or $\operatorname{sigmoid}(u_\mathrm{conv})$ as their posteriors remain close to their uniform priors and therefore provide little additional insight.

%%%%%%%%%%%%%%%%%%%%%%%%%%%%%%%%%%%%%%%%%%%%%%%%%%%%%%%%%%%%%%%

For the transition radii, the posteriors differ substantially from the priors.
The prior sampling allows all normalised transition radii to approach zero and one.
Thus, extreme configurations are included, such as $ r_3/R \approx 1$ and $r_1/R \approx 0$.
However, many of these configurations are rejected by the mass constraint.
The posterior consequently becomes more concentrated and excludes most of these edge cases.

%%%%%%%%%%%%%%%%%%%%%%%%%%%%%%%%%%%%%%%%%%%%%%%%%%%%%%%%%%%%%%%

Only the composition-gradient models depend on the composition parameters.
Their posteriors remain somewhat closer to the priors.
This indicates that the mass constraint places weaker restrictions on the composition parameters and that the allowed composition-gradient models very likely span a broader range than the models shown here.
A completely uniform composition prior would, however, be inefficient.
It would assign substantial probability to physically implausible
profiles.
For example, it could generate iron-rich outer layers above
H--He-rich inner layers.
We therefore deliberately chose priors that roughly follow the ordering found in the purely adiabatic models (H--He $\rightarrow$ water $\rightarrow$ rock $\rightarrow$ iron from the outside to the interior).
This choice provides a more meaningful comparison between the two
model classes and avoids spending most of the sampling effort on
clearly implausible compositions.

%%%%%%%%%%%%%%%%%%%%%%%%%%%%%%%%%%%%%%%%%%%%%%%%%%%%%%%%%%%%%%%

%%%%%%%%%%%%%%%%%%%%%%%%%%%%%%%%%%%%%%%%%%%%%%%%%%%%%%%%%%%%%%%

\section{The impact of atmospheric metallicities}
\label{sec:atmos_metal}

%%%%%%%%%%%%%%%%%%%%%%%%%%%%%%%%%%%%%%%%%%%%%%%%%%%%%%%%%%%%%%%

\begin{figure}

    \centering
    \includegraphics[width=\linewidth]{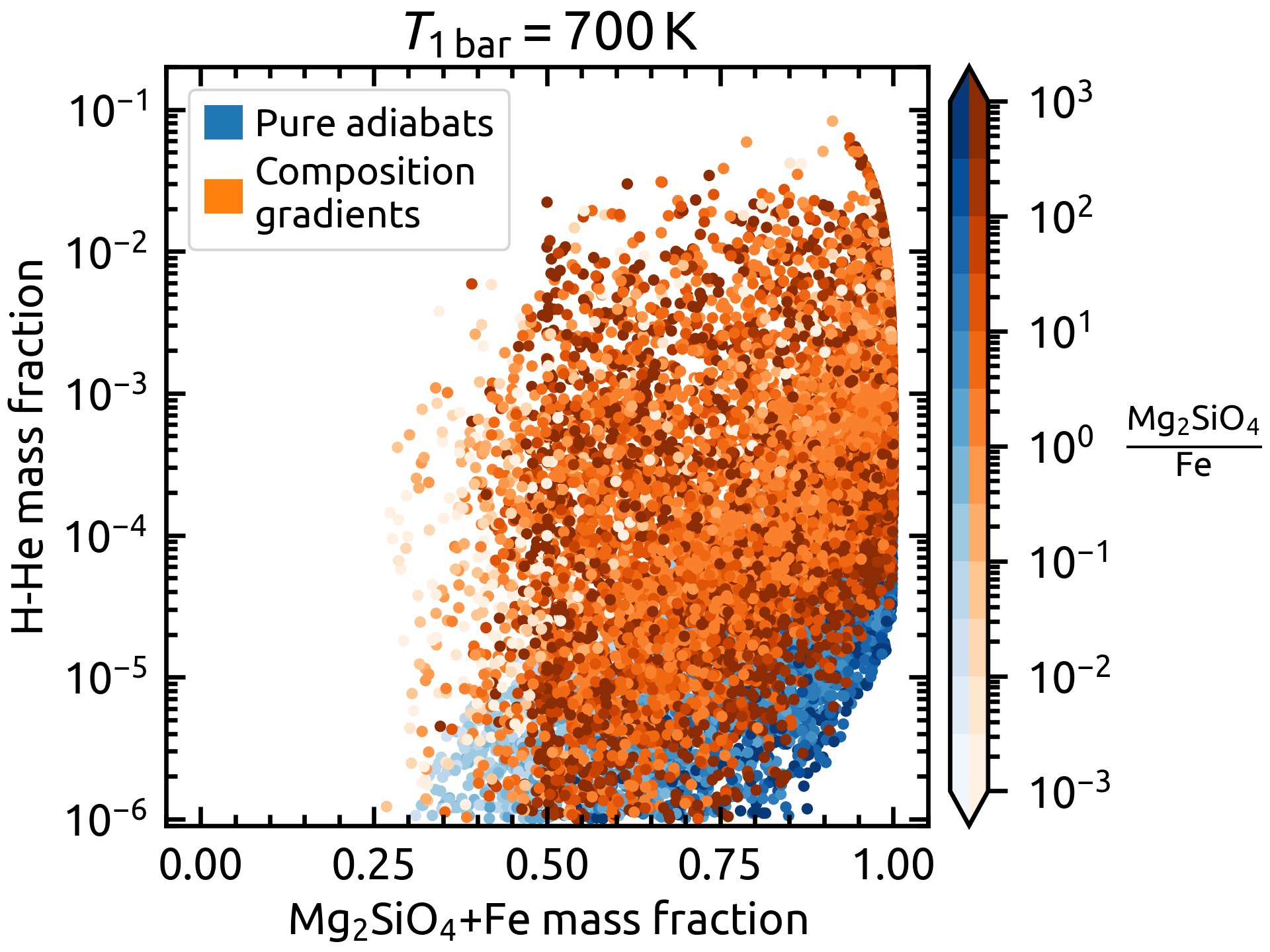}
    \caption{
    Same as Fig.~\ref{fig:K2-18b_hhe-vs-rock_pure+grad-s1-T300_by-Mg2SiO4-Fe} but for TOI-270\,d and the high 1\,bar temperature scenario.
    }
    \label{fig:TOI-270d_hhe-vs-rock_pure+grad-s1-T700_by-Mg2SiO4-Fe}

    \vspace{0.5cm}

    \centering
    \includegraphics[width=\linewidth]{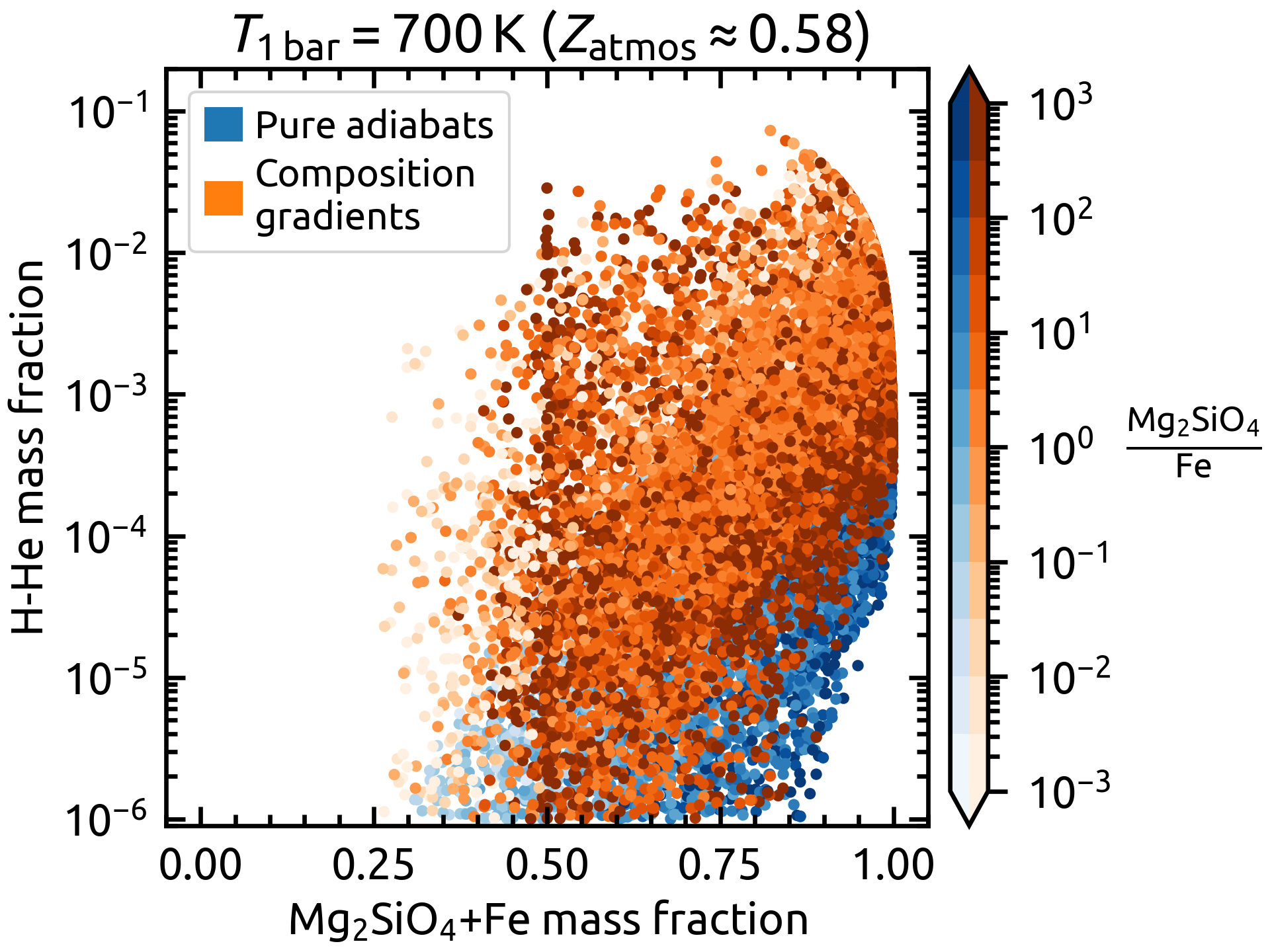}
    \caption{
    Same as Fig.~\ref{fig:K2-18b_hhe-vs-rock_pure+grad-s1-T300_by-Mg2SiO4-Fe} but for TOI-270\,d (the high 1\,bar temperature scenario) and with an enforced atmospheric metallicity of $Z_\mathrm{atmos}\approx0.58$.
    }
    \label{fig:TOI-270d_atmos100_hhe-vs-rock_pure+grad-s1-T700_by-Mg2SiO4-Fe}
\end{figure}

%%%%%%%%%%%%%%%%%%%%%%%%%%%%%%%%%%%%%%%%%%%%%%%%%%%%%%%%%%%%%%%

As mentioned in Sect.~\ref{sec:Methods}, the models we presented in the main text do not impose an atmospheric metallicity constraint in order to keep the solutions as general as possible.
While an atmospheric metallicity measurement can further constrain the models, atmospheric observations typically probe low-pressure regions in the atmosphere while our models begin at 1\,bar. 
It remains unclear whether the composition in the uppermost atmosphere can be linked to deeper regions.
For example, deep radiative zones could prevent efficient mixing between the two \citep{Muller2024, Muller2026}.

%%%%%%%%%%%%%%%%%%%%%%%%%%%%%%%%%%%%%%%%%%%%%%%%%%%%%%%%%%%%%%%

Furthermore, atmosphere models can disagree on the retrieved atmospheric metallicities \citep[see e.g. the difference between][]{Benneke2024, Constantinou2026}. 
In fact, even for Uranus and Neptune, for which we have more data, the atmospheric metallicity remains unknown \citep{Hueso2020,Doucot2026}.

%%%%%%%%%%%%%%%%%%%%%%%%%%%%%%%%%%%%%%%%%%%%%%%%%%%%%%%%%%%%%%%

Finally, the measured atmospheric metallicity cannot be easily translated into a heavy-element mass fraction for interior models. 
It holds that
\begin{equation}
Z_\mathrm{atmos} = \frac{\bar{x}}{(1-Z_\odot)/Z_\odot+\bar{x}}\approx\frac{\bar{x}}{70+\bar{x}}, \label{eq:B1}
\end{equation}
where $Z_\mathrm{atmos}$ denotes the heavy-element mass fraction in the atmosphere \citep{Chachan2026}.
$\bar{x}$ is given by
\begin{equation}
    \bar{x}=\sum_i\left(\frac{Z_i}{Z}\right)_\odot x_i,
    \label{eq:B2}
\end{equation}
where $x_i$ is the factor by which the $i$-th element is enriched compared to solar values.
Atmosphere models in the literature (such as in Table~\ref{tab:planet_data}) often report $\bar{x}$ or individual $x_i$ in their results.

%%%%%%%%%%%%%%%%%%%%%%%%%%%%%%%%%%%%%%%%%%%%%%%%%%%%%%%%%%%%%%%

Equation~\eqref{eq:B1} can only be used consistently if all $x_i$ are known.
But even then, interior models consider only a few representative components that represent the heavy elements. 
For example, our models consider the existence of oxygen in the atmosphere (in the form of water) but no other elements such as carbon or nitrogen.
We hence treated water as a proxy for all metals.
This leaves us with two options to estimate $Z_\mathrm{atmos}$ from Eq.~\eqref{eq:B1}:
one possibility is to treat the Sun in an analogous fashion by setting $Z_{\mathrm{Oxygen},\odot}=Z_\odot$ and $(Z_i)_\odot=0$ for all elements besides oxygen in Eq.~\eqref{eq:B2}.
Alternatively, one can leave all $(Z_i)_\odot$ at their true values and adjust the values of $x_i$.
But regardless of the above choice, the treatment of the atmosphere is no longer chemically consistent with the atmosphere model and yields a different $Z_\mathrm{atmos}$ than the atmosphere model with all heavy elements would obtain.

%%%%%%%%%%%%%%%%%%%%%%%%%%%%%%%%%%%%%%%%%%%%%%%%%%%%%%%%%%%%%%%

Despite this inconsistency, it is still interesting to investigate the effect of this additional constraint on our inferred models.
\cite{Benneke2024} directly report the heavy-element mass fraction $Z_\mathrm{atmos}$ for TOI-270\,d.
Figure~\ref{fig:TOI-270d_atmos100_hhe-vs-rock_pure+grad-s1-T700_by-Mg2SiO4-Fe} shows interior models that are consistent with their reported $Z_\mathrm{atmos}\approx0.58$, which can be directly compared to models where no such constraint is imposed in Fig.~\ref{fig:TOI-270d_hhe-vs-rock_pure+grad-s1-T700_by-Mg2SiO4-Fe}.
Notably, both model types are only moderately affected by the additional atmospheric constraint.
They can simply compensate for the higher heavy-element abundances in the atmosphere by including H--He in deeper regions and hence cover the same region in the solution spaces of Figs.~\ref{fig:TOI-270d_hhe-vs-rock_pure+grad-s1-T700_by-Mg2SiO4-Fe} and \ref{fig:TOI-270d_atmos100_hhe-vs-rock_pure+grad-s1-T700_by-Mg2SiO4-Fe}.
For purely adiabatic models and $T_{1\,\mathrm{bar}}=700\,$K, the Spearman rank correlation coefficients now are $-0.23$, $-0.44$, and $0.70$ for H--He versus H$_2$O/Mg$_2$SiO$_4$, Mg$_2$SiO$_4$/Fe, and Fe correlations, respectively.
For composition-gradient models, the analogous numbers are $-0.55$, $-0.19$, and $0.29$.
These numbers are similar to the numbers depicted in Fig.~\ref{fig:TOI-270d_corr}.
We note, however, that this is only a qualitative test, since the atmosphere of these interior models contains only oxygen in the form of water and is therefore inconsistent with the atmosphere model of \cite{Benneke2024}.
We encourage further research in this direction, which would require the determination of the atmospheric composition in deeper regions, as well as more sophisticated interior models that differentiate among different chemical species and have an advanced atmosphere-interior treatment.

%%%%%%%%%%%%%%%%%%%%%%%%%%%%%%%%%%%%%%%%%%%%%%%%%%%%%%%%%%%%%%%

\end{appendix}

%%%%%%%%%%%%%%%%%%%%%%%%%%%%%%%%%%%%%%%%%%%%%%%%%%%%%%%%%%%%%%%

\end{document}